\pdfoutput=1
\documentclass[a4paper,11pt]{article}

\usepackage{jheppub} 

\usepackage[T1]{fontenc} 
\usepackage{lmodern}
\usepackage{newtxtext,amssymb,amsmath,extarrows,slashed,hyperref}
\usepackage{graphicx,microtype,physics,framed,xcolor}
\usepackage{mdframed,adjustbox} 
\newmdenv[backgroundcolor=gray!15,%
skipabove=0pt,%
skipbelow=5pt,%
leftmargin=0pt,%
rightmargin=0pt,%
innertopmargin=-5pt,%
innerbottommargin=7pt,%
innerleftmargin=2pt,%
innerrightmargin=2pt,%
splittopskip=0pt,%
splitbottomskip=0pt,%
linewidth=0pt,%
nobreak=true]%
{keyeqn}

\graphicspath{{Fig/}}

\makeatletter
\gdef\@fpheader{} 
\makeatother

\title{\boldmath  BCS in the Sky: Signatures of Inflationary Fermion Condensation}

\author[a,b,c]{Xi Tong, }
\author[b,c]{Yi Wang, }
\author[b,c]{Chen Zhang}
\author[d,b,c]{and Yuhang Zhu}

\affiliation[a]{Department of Applied Mathematics and Theoretical Physics, University of Cambridge,\\Wilberforce Road, Cambridge, CB3 0WA, UK}
\affiliation[b]{Department of Physics, The Hong Kong University of Science and Technology,\\Clear Water Bay, Kowloon, Hong Kong, P.R. China}
\affiliation[c]{The HKUST Jockey Club Institute for Advanced Study,\\The Hong Kong University of Science and Technology,\\Clear Water Bay, Kowloon, Hong Kong, P.R. China}
\affiliation[d]{Cosmology, Gravity and Astroparticle Physics Group, Center for Theoretical Physics of the Universe,\\
Institute for Basic Science, Daejeon 34126, Korea}

\emailAdd{xt246@cam.ac.uk}
\emailAdd{phyw@ust.hk}
\emailAdd{iasczhang@ust.hk}
\emailAdd{yhzhu@ibs.re.kr}

\abstract{We consider a Bardeen-Cooper-Schrieffer (BCS)-like model in the inflationary background. We show that with an axial chemical potential, the attractive quartic fermion self-interaction can lead to a BCS-like condensation. In the rigid-de Sitter (dS) limit of inflation where backreaction from the inflaton and graviton is neglected, we perform the first computation of the non-perturbative effective potential that includes the full spacetime curvature effects in the presence of the chemical potential, subject to the mean-field approximation whose validity has been checked via the Ginzburg criterion. The corresponding BCS phase transition is always first-order, when the varying Hubble is interpreted as an effective Gibbons-Hawking temperature of dS spacetime. In the condensed phase, the theory can be understood from UV and IR sides as fermionic and bosonic, respectively. This leads to distinctive signatures in the primordial non-Gaussianity of curvature perturbations. Namely, the oscillatory cosmological collider signal is smoothly turned off at a finite momentum ratio, since different momentum ratios effectively probe different energy scales. In addition, such BCS phase transitions can also source stochastic gravitational waves, which are feasible for future experiments.}

\begin{document}
\maketitle
\flushbottom


\section{Introduction}
The phenomenon of superconductivity is one of the most important discoveries in modern physics. In 1911, H. Onnes first observed that the mercury electrical resistance drops to zero below 4K, an effect he called superconductivity \cite{Onnes}. Since then, many insightful attempts have been made to give a theoretical understanding of superconductivity, notably the phenomenological London equations in 1935 \cite{London} and the Landau-Ginzburg model in 1950 \cite{Ginzburg}. The successful microscopic theory of superconductors was finally put forward by J. Bardeen, L.N. Cooper and J.R. Schrieffer in 1957. In the Bardeen-Cooper-Schrieffer (BCS) theory \cite{Bardeen:1957mv}, the electrons moving in the ionic lattice can develop an attractive interaction by exchanging phonons. At low temperatures, this attractive force binds electrons into Cooper pairs. These bosonic Cooper pairs then condense into the BCS vacuum, spontaneously breaking electromagnetic $U(1)_{EM}$ symmetry, introducing the Meissner effect and a non-zero supercurrent.

At the core of the BCS theory is the instability of electrons near the Fermi surface. Namely, as long as the interaction is attractive, and the fermion number is supported by a finite chemical potential, the Fermi sea is unstable against the formation of Cooper pairs, no matter how weak the attractive interaction is. The universality of the BCS model assumption makes it a flexible framework to generalize to other contexts. Over the past decades, the BCS theory and its variations played crucial roles in many aspects of modern physics from condensed matter to fundamental high energy theory. To name a few, BCS-like models are used for QCD chiral dynamics~\cite{Nambu:1961fr, Buballa:2003qv,Osipov:2013fka, Holdom:2017gdc} and color superconductivity~\cite{Rajagopal:2000wf,Alford:1997zt, Alford:2007xm}, early universe model building \cite{Giacosa:2008rw,Alexander:2008vt}, neutrino condensation~\cite{Caldi:1999db}, dark energy/dark matter candidate \cite{Alexander:2009uu,Alexander:2008zf,Inagaki:2003qq,Alexander:2016glq,Alexander:2018fjp,MacKenzie:2020scl,MacKenzie:2021ghe,Garani:2022quc}, and fermion dynamics around black holes~\cite{Hartman:2010fk}~.

Cosmology, on the other hand, is in a sense very much akin to condensed matter physics. For instance, Lorentz symmetry is always spontaneously broken (by the lattice in condensed matter systems and by the cosmic expansion in cosmology). There is, in both cases, a separation of a non-trivial homogeneous background and inhomogeneous perturbations. The observables both entail correlation functions of perturbations. In fact, both condensed matter systems and cosmology can be classified according to a common set of symmetry breaking patterns \cite{Nicolis:2015sra}. Such a link between the two systems motivates us to look for the analogy of BCS condensation in a cosmological setup.

For instance, let us consider the fermion dynamics during inflation. It has been shown that the rolling inflaton can naturally provide an external chemical potential that supports a number of fermions \cite{Adshead:2015kza,Chen:2018xck,Adshead:2018oaa,Hook:2019zxa,Sou:2021juh,Chen:2023txq}. Then after turning on a weak attractive interaction, the fermions may pair up and condense, forming a BCS-like state. The crucial difference here is, of course, that the universe is expanding and the BCS mechanism must be modified by the spacetime curvature effects. We will see that the spacetime curvature can be effectively considered as a thermal effect, with a temperature set by the Hubble parameter during inflation. From the experience of condensed matter systems, we know that superconductivity typically requires low temperature. If the spacetime expansion is too fast, the effective temperature is too high, and the Cooper pairs may be teared apart, breaking the condensate. It is then possible to find a ``critical temperature'' for an inflationary BCS system. We also note that in the previous analysis of BCS in cosmology~\cite{Candelas:1975du,Elizalde:1994fh,Elizalde:1995bm,Elizalde:1995kg,Inagaki:1995jp,Inagaki:1997kz,Alexander:2006we,Addazi:2017qus, Giacosa:2008rw, Caldi:1999db,Alexander:2008vt,Alexander:2009uu}, the curvature effect has not yet been fully studied, and only leading-order results in Hubble constant expansion have been achieved so far~\cite{Elizalde:1994fh,Elizalde:1995bm,Alexander:2009uu}. In the dS-invariant case, all-order results have been achieved but in order to generate a non-zero gap, either one needs to go to lower-dimensional spacetime \cite{Elizalde:1995kg,Inagaki:1995jp} or introduce super-Planckian physics \cite{Giacosa:2008rw}. It is thus doubtful whether any exponentially small gap can truly exist in an expanding spacetime.

To answer this theoretical question and explore the phenomenology of BCS in the sky, we set out in this paper to compute the effective action of a class of BCS-like models during inflation. We find an analytical result for the effective potential exact to all orders in the Hubble expansion. Inspecting this effective potential, we show that the BCS phase transition is always first-order under the naturalness assumption, suggesting the removal of any exponentially small gap by cosmic expansion.	Since any expanding flat FRW universe can be locally approximated as dS when the Hubble parameter is slowly varying, our analysis can be thought of as an approximate IR completion of BCS in any FRW spacetime. In addition, we show that the squeezed limit of primordial bispectrum probes different phases of the BCS theory, leading to distinctive cosmological collider \cite{Chen:2009zp,Baumann:2011nk,Noumi:2012vr,Arkani-Hamed:2015bza} signals that are feasible for future non-Gaussianity surveys \cite{SimonsObservatory:2018koc,CMB-S4:2016ple,NASAPICO:2019thw,Amendola:2016saw,DESI:2016fyo,Dore:2014cca,SKA:2018ckk}. If the phase transition happens during inflation, the bubble collision process gives rise to stochastic gravitational waves that can be probed by a variety of detection methods, such as SKA~\cite{Janssen:2014dka}, LISA~\cite{LISA:2017pwj} and DECIGO~\cite{Kawamura:2006up}.

This paper is structured as follows. In Sect.~\ref{sec2}, we start with a brief review of BCS in the LAB and try to generalize it to a cosmological model, while giving some qualitative understandings. In Sect.~\ref{mainSect}, we quantitatively analyze our BCS model first in flat spacetime and then in dS. After computing the non-perturbative effective potential, we examine the phase transition properties in Sect.~\ref{phaseTransitionSect}. The impact on observables is discussed in Sect.~\ref{ObsSect}. At last, we conclude and give future prospects in Sect.~\ref{ConclusionSect}.

\paragraph{Notations and conventions}
Throughout the paper, we use the $(-,+,+,+)$ metric sign convention. The Pauli matrices with Lorentz indices are defined as $\sigma^\mu=(1,\boldsymbol{\sigma})$, $\bar{\sigma}^\mu=(1,-\boldsymbol{\sigma})$, where $\boldsymbol{\sigma}$ are the usual three Pauli matrices. The Dirac matrices and its algebra are given by
\begin{align}
	\gamma^\mu=\left(\begin{array}{cc}
		0 & \sigma^\mu\\
		\bar{\sigma}^\mu & 0
	\end{array}\right)\quad,\quad\gamma^5=\left(\begin{array}{cc}
	-1 & ~0\\
	~~0 & ~1
\end{array}\right)\quad,\quad \{\gamma^\mu,\gamma^\nu\}=-2\eta^{\mu\nu}\quad,\quad \{\gamma^5,\gamma^\mu\}=0~.
\end{align}
We will also adopt the Van der Waerden notation for two-component spinors and the conventions follow from \cite{Martin:1997ns} and \cite{Dreiner:2008tw}. In the dS context, we use dimensionless tilde variables for any parameters measured in Hubble units, for example, $\tilde{\kappa}\equiv \kappa/H$, $\tilde{\Delta}\equiv \Delta/H$, $\tilde{\mu}\equiv \mu/H$, etc. A prime on quantum average values denotes the momentum $\delta$-function stripped average value,
\begin{align}
	\Big\langle\mathcal{O}(\{\mathbf{k}\})\Big\rangle\equiv (2\pi)^3\delta^3\left(\sum_a \mathbf{k}_a\right)\Big\langle\mathcal{O}(\{\mathbf{k}\})\Big\rangle'~.
\end{align}


\section{From superconductors to the universe}\label{sec2}

   In the original BCS theory, the conducting-band electrons move in an ionic lattice and couple to the quanta of lattice vibration known as the phonon. In the zero temperature and decoupling limit, the electrons fill up the lowest energy states and form a sharp Fermi sphere. However, the coupling to the phonon introduces a non-local attractive interaction for pairs of electrons with opposite spin ($s$-wave pairing channel). Consequently, it becomes energetically favorable for the electrons to pair up and redistribute themselves, causing the spreading of the Fermi sphere. The width of the spread is given as the gap:
		\begin{align}
	\Delta_{\text{BCS}}\approx 2\omega_D e^{-\frac{1}{g N}}~,\label{BCSgap}
\end{align}
where $\omega_D$ is the Debye frequency, $g$ is the coupling strength and $N$ is the density of states near the Fermi surface. The BCS ground state is then a two-mode squeezed state
\begin{align}
	|\text{BCS}\rangle=\prod_{\mathbf{k}} \left(u_k+v_k e^{i\theta_k}c_{\mathbf{k},\uparrow}^\dagger c_{-\mathbf{k},\downarrow}^\dagger\right)|0\rangle~,\label{BCSgroundState}
\end{align}
where $\uparrow\downarrow$ denotes the spin-z eigenvalue, not the helicity. Any fermionic excitation on this ground state is gapped by $\Delta$ in the energy spectrum. Since the electrons carry electromagnetic charge, this condensate of electrons necessarily breaks $U(1)_{EM}$ and leads to superconductivity.

In contrast, in a cosmological setup, the BCS model structure must be slightly modified. This is because the universe must stay electrically neutral in the absence of any positively charged lattice. Therefore it is not possible to have a condensate of fermions charged under $U(1)_{EM}$ \footnote{It is possible, however, to consider condensates that are charged under some dark $U(1)$ symmetry or spontaneously broken non-Abelian gauge symmetry. However, we prefer to focus on the simplest scenario and choose not to introduce gauge symmetries in this work.}. Instead, we require fermions and anti-fermions to pair up. Consider a massless Dirac fermion model,
\begin{align}
	\mathcal{L}_0=i\bar{\psi}\slashed{\nabla}\psi~,
\end{align}	
where $\slashed{\nabla}=\gamma^\mu \nabla_\mu$ is the covariant derivative with the spacetime spin connection. This fermion may be understood as a new degree of freedom active during inflation, for example those from Grand Unification model building. Now the global $U(1)_V$ symmetry is exact, and the vector fermion current $J_V^\mu=J_R^\mu+J_L^\mu$ is conserved, where $J_R^\mu$ and $J_L^\mu$ are the right- and left-handed fermion current. The global $U(1)_A$ symmetry, however, may not be exact. This corresponds to the non-conservation of the axial current, $J_A^\mu=J_R^\mu-J_L^\mu$. Namely, there will be an asymmetry in the number of Cooper pairs with different constituent fermion helicities. To achieve this, we can introduce a parity-odd chemical potential via the operator
\begin{align}
	\mathcal{L}_c=-\kappa J_A^0=-\kappa\bar{\psi}\gamma^0\gamma^5\psi~,
\end{align}
where $\kappa$ is the axial chemical potential. This non-zero chemical potential $\kappa$ supports a net axial charge,
\begin{align}
	Q_A=\int_\Sigma d^3x_\mu J_A^\mu=N_{+}+\bar{N}_{+}-N_{-}-\bar{N}_{-}\neq 0~,\label{axialCharge}
\end{align}
while keeping a neutral vector charge,
\begin{align}
	Q_V=\int_\Sigma d^3x_\mu J_V^\mu=N_{+}+N_{-}-\bar{N}_{+}-\bar{N}_{-}=0~.\label{vectorCharge}
\end{align}
Here $+,-$ denotes the helicity while $N,\bar{N}$ denotes the fermion and anti-fermion, respectively. In the absence of any helical chemical potential \cite{Ambrus:2019ayb,Chernodub:2020yaf,Ambrus:2019khr}, the net helical charge is also zero, we have
\begin{align}
	Q_H=N_{+}-N_{-}-\bar{N}_{+}+\bar{N}_{-}=0~.\label{helicalCharge}
\end{align}
Combining (\ref{axialCharge}), (\ref{vectorCharge}) and (\ref{helicalCharge}), we see that
\begin{align}
	N_{+}=\bar{N}_{+}\neq N_{-}=\bar{N}_{-}~.
\end{align}
Thus our system contains an equal number of fermions and anti-fermions with the same helicity, but there is an imbalance of particle numbers for different helicities. 

Note that another technical reason for choosing an axial chemical potential as opposed to a vector chemical potential is that fermion chemical potential does not produce particles dynamically during inflation, and can be rotated away as shown in \cite{Wang:2019gbi}. In other words, a net vector fermion number density is not preserved by inflationary expansion, and if such a state is prepared initially, the inflationary expansion will dilute the vector fermion number away and drive the state toward the dS-invariant attractor solution of a Bunch-Davies vacuum. In contrast, an axial chemical potential is compatible with Bunch-Davies vacuum and cannot be rotated away when the condensation forms. It can assist dynamical particle production in a dS-invariant way, keeping a constant net axial fermion number density when the production rate and dilution rate achieve a balance.

\begin{figure}[htp] 
	\centering
	\includegraphics[width=9cm]{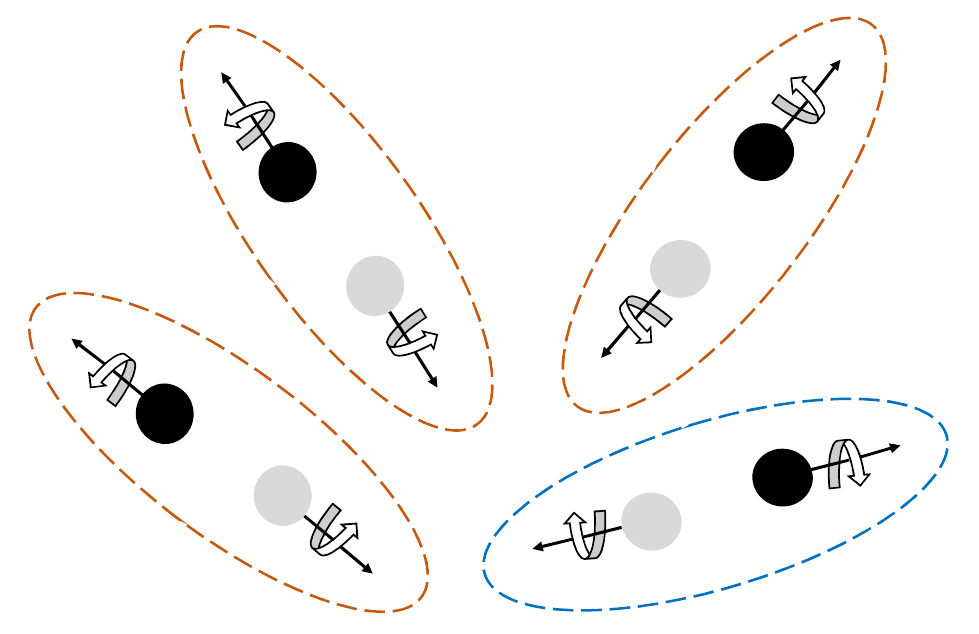}\label{CooperPairCartoon}
	\caption{A cartoon illustration of the Cooper pair constituents in our BCS-like model. Black and gray circles represent fermions and anti-fermions, with their momenta indicated by the arrow and spin labeled with respect to the momenta. Each Cooper pair consists of a fermion-anti-fermion pair with opposite momenta and same helicity. Due to the axial chemical potential, there are more Cooper pairs built from left-handed fermions (red dashed circle) than that from right-handed fermions (blue dashed circle).}
\end{figure}

We are now ready to introduce the attractive self-interaction that gives rise to BCS condensation. The simplest four-fermion interaction that binds fermions with anti-fermions is given by
\begin{align}
	\mathcal{L}_{int}=+\frac{(\bar{\psi}\psi)^2}{M^2}~.\label{4fint}
\end{align}
The plus sign in the front guarantees the attractive nature of the interaction, since having a larger fermionic field bilinear \textit{decreases} the potential energy. Thus fermion fields tend to be drawn closer. Such an interaction can originate from the exchange of a heavy scalar particle of mass scale $M$, or from the $s$-wave channel of the torsion-induced four-fermion interaction \cite{Perez:2005pm,Freidel:2005sn}. Since the energy scale of inflation is not far below that of Grand Unified Theories (GUT) such as the Pati-Salam model \cite{Pati:1974yy,Hartmann:2014fya,Pallis:2012zd,Bryant:2016tzg,Holdom:2014hla}, one can treat the four-fermion term as EFT operators of a broken GUT group, in a similar way that these operators appear in the SMEFT~\cite{Cepedello:2022pyx}. For simplicity, we shall first focus on (\ref{4fint}) and comment on other types of interaction in Sect.~\ref{U1A}. We also note that different from the non-local interaction of electrons in real BCS systems, the quartic self-interaction here appears to be local. Of course from an EFT perspective, local/non-local is a relative notion depending on the hierarchy between the energy scale of the probe and the underlying mechanism. The UV completion of $\mathcal{L}_{int}$ can actually be a non-local exchange of massive scalars at the scale of $M$. This attractive interaction responsible for condensation is then an effective description at the gap energy scale $\Delta$. As long as $\Delta\ll M$, one can integrate out the heavy meson, and the non-local UV picture can be reduced to a local vertex in the IR. After introducing this self-interaction, one expects the fermions and anti-fermions of the same helicity will form Cooper pairs in the $s$-wave channel (see FIG.~\ref{CooperPairCartoon} for a cartoon depiction of the physical picture). Now although the helicity imbalance is still present, it becomes hidden in the internal structure of Cooper pairs, and is not visible in the IR picture at leading order. This fact will affect the parity property of the resulting cosmological collider signals, as we will demonstrate later.

In summary, our BCS Lagrangian reads
\begin{align}
	\nonumber S[\psi]&=\int d^4x\sqrt{-g}\left(\mathcal{L}_0+\mathcal{L}_c+\mathcal{L}_{int}\right)\\
	&=\int d^4x\sqrt{-g}\left[i\bar{\psi}\slashed{\nabla}\psi-\kappa\bar{\psi}\gamma^0\gamma^5\psi+\frac{(\bar{\psi}\psi)^2}{M^2}\right]~,
\end{align}
where $g_{\mu\nu}=a(\tau)^2\eta_{\mu\nu}$ is the background spacetime metric and $\mathcal{D}_\mu$ is the covariant derivative. For clarity, we will analyze our BCS model progressively, first in flat spacetime and then in dS and actual inflation.


\section{BCS in the sky}\label{mainSect}

\subsection{Flat spacetime case}\label{flatSpacetimeSect}
 Here, we review the case of flat spacetime that has already been well studied, setting the footage to the dS case.
 
  Reducing the covariant derivative to normal derivatives, the Lagrangian reads
\begin{align}
   	\mathcal{L}_{\rm{flat}}=i\bar{\psi}\slashed{\partial}\psi-\kappa\bar{\psi}\gamma^0\gamma^5\psi+\frac{(\bar{\psi}\psi)^2}{M^2}~.
\end{align}
In the full theory partition function, the four-fermion interaction term can be resolved to a quadratic interaction by introducing an auxiliary field $\Delta$,
\begin{align}
\nonumber Z&=\int\mathcal{D}\psi\mathcal{D}\bar{\psi}\,e^{i\int d^4 x\mathcal{L}_{\rm{flat}}}\\
&=\int\mathcal{D}\psi\mathcal{D}\bar{\psi}\mathcal{D}\Delta\,e^{i\int d^4 x \bar{\psi}(i\slashed{\partial}-\kappa\gamma^0\gamma^5+\Delta)\psi-\frac{M^2\Delta^2}{4}}\nonumber\\
 &\equiv\int\mathcal{D}\Delta\, e^{-i\int d^4x \frac{M^2 \Delta^2}{4}}Z_f[\Delta]~,
\end{align}
where we have introduced
\begin{align}
   	Z_f[\Delta]=\int\mathcal{D}\psi\mathcal{D}\bar{\psi}\,e^{i\int d^4x \bar{\psi}(i\slashed{\partial}-\kappa\gamma^0\gamma^5+\Delta)\psi}
\end{align}
as the partition function for the fermion field with ``effective mass'' $\Delta$. Then the full partition function can be written as a bosonic path integral weighed by an effective action of the auxiliary field $\Delta$,
\begin{align}
   	Z=\int \mathcal{D}\Delta e^{i S_{\text{eff}}[\Delta]}\quad,\quad S_{\text{eff}}\equiv-\int d^4 x\frac{M^2\Delta^2}{4}-i \ln Z_f[\Delta]~.\label{SeffFlatDef}
\end{align}
Notice that this effective theory of the bosonic field $\Delta$ should be understood as the IR face of the original UV theory of interacting fermions.

To determine the vacuum state of the effective theory in the IR, it suffices to take the mean-field description of the auxiliary field $\Delta(x)=\Delta\approx \mathrm{const}$ and integrate out the fermions, so that the partition function can be easily computed as an infinite determinant as
\begin{align}
Z_f[\Delta]&={\rm{Det}}{\left(i\slashed{\partial}-\kappa\gamma^0\gamma^5+\Delta\right)}=\prod_k \det
\begin{pmatrix}
\Delta&\quad\sigma^\mu k_\mu-\kappa\\
\bar{\sigma}^\mu k_\mu+\kappa&\Delta
\end{pmatrix}~,
\end{align} 
or
\begin{align}
\ln{Z_f}=\int d^4x\int\frac{d^4k}{(2\pi)^4}\ln\left[\Delta^2-\omega^2+(|\mathbf{k}|-\kappa)^2\right]\left[\Delta^2-\omega^2+(|\mathbf{k}|+\kappa)^2\right]~.\label{Zpsi}
\end{align}
Finishing the loop momentum integral in (\ref{Zpsi}) and inserting it back to (\ref{SeffFlatDef}), we obtain the effective potential defined as the leading order effective action in the uniform limit,
\begin{align}
   	S_{\text{eff}}=\int d^4x\left[-V_{\text{eff}}(\Delta)+\mathcal{O}(\partial^2)\right]~,
\end{align}
with
\begin{align}
V_{\text{eff}}(\Delta)=\frac{1}{4}\left[M^2-\frac{\Lambda^2}{\pi^2}+\frac{2\kappa^2(1-\ln 2)}{\pi^2}\right]\Delta^2+\frac{\kappa^2}{2\pi^2}\Delta^2\ln\frac{\Delta}{\Lambda}-\frac{\Delta^4}{8\pi^2}\left(\ln\frac{\Delta}{\Lambda}+\frac{1}{4}-\ln 2\right)~.\label{flatV}
\end{align}
Here $\Lambda$ is an explicit loop momentum cutoff, which should be replaced by renormalized parameters and scales after renormalization. 
However, we are going to consider embedding the BCS model in a time-dependent cosmological set-up later, and renormalization in a time-dependent system (such as cosmology) is by no means a simple task. For example, it is not clear to us what renormalization conditions should be posed to eliminate the cutoffs ($\Lambda$) and bare variables ($M$), nor at what time should they be imposed. One possible choice is to fix the second- and fourth-order derivative at the gap:
\begin{align}
	\frac{\partial^2 V_{\mathrm{eff}}}{\partial\Delta^2}\Big|_{\Delta=\Delta_0}=\mu^2~,\quad \frac{\partial^4 V_{\mathrm{eff}}}{\partial\Delta^4}\Big|_{\Delta=\Delta_0}=\lambda~,\label{exampleRenormCond}
\end{align}
where $\mu,\lambda$ are some constants to be measured by experiments. Of course, the question in cosmology is that unlike in particle scatterings where pole masses and threshold couplings are known from experiments, there is \textit{no} such measurement that tells us the value of $\mu,\lambda$ at $\Delta_0$. In addition, some parameters entering the renormalization conditions are time-dependent, so it is not useful to stick to a fixed condition like (\ref{exampleRenormCond}). As a result, in favor of practicality, we do not pose any explicit renormalization conditions, but understand the cutoffs $\Lambda,M$ as \textit{defining} the renormalization conditions. Namely, we trade the functional dependence by
\begin{align}
	M=f_1(\mu,\lambda)~,\quad	\Lambda = f_2(\mu,\lambda)~,
\end{align}
and study the effective potential by varying $M,\Lambda$ as an alternative to $\mu,\lambda$. We also remark that the effective potential $V_{\mathrm{eff}}$ is formulated as a tool to determine the ground state of the system, which is why we care about the functional dependence on the gap $\Delta$. However, just like the Coleman-Weinberg effective potential, $V_{\mathrm{eff}}$ is always defined up to a $\Delta$-independent $\mathcal{O}(\kappa^4)$ constant. This $\mathcal{O}(\kappa^4)$ constant remainder does not affect the determination of the gap, but can be understood to represent the free energy of the free relativistic Fermi gas.

The lowest energy state of the theory is determined by the minimum of the effective potential. This gives the gap equation,
\begin{align}
   	\frac{\partial V_{\rm{eff}}}{\partial \Delta}\Big|_{\Delta=\Delta_0}=0~.\label{flatGapEq}
\end{align}
The location of the minimum $\Delta=\Delta_0$ corresponds to the gap of BCS condensation. A way to see this is through computing the average value of the fermion bilinear in the UV theory of fermions, it is equal to the average value of the gap field in the IR theory of bosons:
\begin{align}
	\nonumber\langle\bar{\psi}\psi(x)\rangle_{UV}&=\int \mathcal{D}\psi\mathcal{D}\bar{\psi}e^{i\int d^4 x'\mathcal{L}_{\rm{flat}}}\bar{\psi}\psi(x)\\
	\nonumber&=\int \mathcal{D}\Delta e^{-i\int d^4x' \frac{M^2 \Delta^2}{4}}\frac{\delta }{i\delta \Delta(x)}Z_f[\Delta]\\
	\nonumber&=-\int \mathcal{D}\Delta \left(\frac{\delta }{i\delta \Delta(x)}e^{-i\int d^4x' \frac{M^2 \Delta^2}{4}}\right) Z_f[\Delta]\\
	\nonumber&=\frac{M^2}{2}\int \mathcal{D}\Delta e^{i S_{\text{eff}}[\Delta]} \Delta(x)\\
	&=\frac{M^2}{2}\langle\Delta(x)\rangle_{IR}~,
\end{align}
where in the third step we have integrated by parts. For this reason, we call the auxiliary field $\Delta$ the gap field, and regard it as the field of Cooper pairs, which is the analogy of BCS ground state (\ref{BCSgroundState}). The $k\neq 0$ modes are understood as bosonic excitations of the ground state. We will discuss these perturbations and their interesting phenomenology later in Sect.~\ref{ObsSect}.

To solve for the gap, we use the approximation $\Delta_0\ll M,\Lambda,\kappa$, then the gap equation (\ref{flatGapEq}) yields a very similar result as the BCS gap (compare with (\ref{BCSgap})),
\begin{align}
\Delta_0=2\Lambda\,e^{-\frac{\pi^2\bar{M}^2}{2\kappa^2}},\quad\text{with}\quad\bar{M}^2\equiv M^2-\frac{\Lambda^2}{\pi^2}+\frac{3\kappa^2}{\pi^2}~.\label{flatBCSgap}
\end{align}
Naturalness requires $\bar{M}\sim M$ are of a similar magnitude. Perturbative unitarity of the four-fermion interaction also yields the condition $\Lambda\lesssim \pi M$. In general, the chemical potential must also be below the cutoff, hence the hierarchy $\kappa<\Lambda\lesssim \pi M$.

As apparently seen from (\ref{flatBCSgap}), the gap decreases with decreasing chemical potential and weaker coupling. Still, the gap equation \textit{always} admits an exponentially small solution. In other words, condensation can appear no matter how small the chemical potential is or how weak the coupling is. This is, of course, an artifact of our zero temperature assumption. At a finite temperature, we expect thermal fluctuations to destabilize the condensation and unbound the fermions, leading to the disappearance of the gap~\cite{coleman_2015}. In cosmology, in addition to thermal effects present in the early universe, the expansion of spacetime is another important factor. It may thus be dangerous to rely on flat spacetime BCS computations in an expanding universe. Namely, the spacetime curvature may easily remove the exponentially small gap, unbinding the Cooper pairs. To fully understand the BCS condensation in an expanding universe such as inflation, we need to perform an ``IR completion'' of the BCS model regarding the effects of spacetime curvature.


\subsection{Going to dS}

Now let us move on to dS spacetime. The background geometry is given by
\begin{align}
	ds^2=-dt^2+e^{2Ht} d\mathbf{x}^2=a^2(\tau)\left(-d\tau^2+d\mathbf{x}^2\right)\quad,\quad a(\tau)=-\frac{1}{H\tau}~,
\end{align}
where $-\infty<t<\infty$ is the cosmic time and $-\infty<\tau<0$ is the conformal time. Note that spacetime expansion is more transparent in the FRW coordinates $(t,\mathbf{x})$, but the conformal coordinates $(\tau,\mathbf{x})$ are technically cleaner. 
Therefore, we will mostly work with the conformal coordinates in the computations below. An empty dS spacetime is maximally symmetric with 10 isometries. However, to implement BCS condensation, the 3 dS boost symmetries must be broken by the chemical potential. Hence, we are left with 7 symmetries, including 3 spatial translations, 3 rotations and 1 dilation: $(\tau,\mathbf{x})\to (\lambda\tau,\lambda\mathbf{x})$. These are the symmetries that constrain the form of our effective action, as we will see later.

We start by spelling out the covariant Lagrangian in the conformal coordinates:
\begin{align}
	S_{\text{dS}}[\psi]=\int d^4x\sqrt{-g}\left[i\bar{\psi}\gamma^a e^\mu_{~~a}\nabla_\mu\psi-\frac{\partial_{\mu}\phi}{\Lambda_c}e^{\mu}_{~~a}\bar{\psi}\gamma^a\gamma^5\psi+\frac{(\bar{\psi}\psi)^2}{M^2}\right]~.\label{dSMainModel}
\end{align}
Here the covariant derivative is $\nabla_\mu\equiv\partial_\mu+\frac{i}{4}\omega_{\mu ab}\sigma^{ab}$, with the spin connection $\omega_{\mu a b}=-e^\nu_{~~b}\nabla_{\mu}e_{\nu}^{~~c}\eta_{ac}$. The dS vierbein can be chosen to be simply $e_{\mu}^{~~a}=a\delta^a_\mu, e^\mu_{~~a}=a^{-1}\delta^\mu_a$~. The chemical potential is naturally introduced through the dimension-5 coupling to a rolling scalar background $\phi=\phi_0(t)$ (usually the inflaton during inflation). $\Lambda_c$ is an new physics scale arising from inflaton-fermion dynamics and is independent of the scale $M$ of fermion self-interactions. It characterizes the strength of the Yukawa-like coupling between the inflaton and the fermion. Then the rolling speed of the scalar $\frac{\partial_{\mu}\phi}{\Lambda_c}=a(\tau) \kappa\delta^0_\mu $ provides the chemical potential,
\begin{align}
	\kappa\equiv \frac{\dot{\phi}_0}{\Lambda_c}~,
\end{align}
where $\dot{\phi}_0\equiv \partial_t \phi_0\approx \mathrm{const}$. The size of the chemical potential is bounded by $\kappa^2<\dot{\phi}_0<\Lambda_c^2$ due to perturbative unitarity \cite{Chen:2018xck,Wang:2019gbi}.  Following the treatment in the preceding section, we introduce the gap field to resolve the quartic self-interaction, and define an effective action for the gap field by integrating out the fermions,
\begin{align}
	Z=\int \mathcal{D}\psi\mathcal{D}\bar\psi\, e^{i S_{\text{dS}}[\psi]}=\int \mathcal{D}\Delta e^{i S_{\text{eff}}[\Delta]}~,
\end{align}
with
\begin{align}
	S_{\text{eff}}[\Delta]=-\int d^4 x \sqrt{-g} \frac{M^2\Delta^2}{4}-i \ln Z_f[\Delta]~,
\end{align}
and
\begin{align}
Z_f[\Delta]=\int\mathcal{D}\psi\mathcal{D}\bar\psi\, e^{i S_f[\psi]}~,~
S_f[\psi]=\int d^4x\sqrt{-g}~\bar{\psi}\left(i\gamma^a e^\mu_{~~a}\nabla_\mu-\kappa\gamma^0\gamma^5+\Delta\right)\psi~.\label{SkappaDelta}
\end{align}

Unlike in flat spacetime, tracing over the fermion spectrum in dS is more non-trivial. We need to inspect its dynamics more clearly before performing the path integral over $\psi$. Due to the loss of time translational symmetry and the dS boosts, it is technically more convenient to work with the two-component-Weyl-spinor representation of the Dirac field,
\begin{align}
	\psi=
	\begin{pmatrix}
		\zeta\\
		\chi^\dag
	\end{pmatrix}~.
\end{align}
In the mean-field limit, $\Delta\approx\mathrm{const}$, the fermions can be quantized and expanded into orthogonal modes with different spatial momenta and helicities,
\begin{align}
	&\zeta_\alpha\left(\tau,\mathbf{x}\right)=\int\frac{d^3k}{(2\pi)^3}\sum_{s=\pm}\left[h^s_\alpha(\hat{\mathbf{k}})u_s(\tau,k)b^s_{\mathbf{k}}e^{i\mathbf{k}\cdot\mathbf{x}}+h^{s\dagger}_{\dot{\gamma}}(\hat{\mathbf{k}})\bar{\sigma}^{0\dot{\gamma}\beta}\epsilon_{\alpha\beta}v^*_s(\tau,k)d^s_{\mathbf{k}}{}^\dagger e^{-i\mathbf{k}\cdot\mathbf{x}} \right]~,\\
	&\chi_\alpha\left(\tau,\mathbf{x}\right)=\int\frac{d^3k}{(2\pi)^3}\sum_{s=\pm}\left[h^s_\alpha(\hat{\mathbf{k}})u_s(\tau,k)d^s_{\mathbf{k}}e^{i\mathbf{k}\cdot\mathbf{x}}+h^{s\dagger}_{\dot{\gamma}}(\hat{\mathbf{k}})\bar{\sigma}^{0\dot{\gamma}\beta}\epsilon_{\alpha\beta}v^*_s(\tau,k)b^s_{\mathbf{k}}{}^\dagger e^{-i\mathbf{k}\cdot\mathbf{x}} \right]~,
\end{align}
with the spinor-helicity basis
\begin{align}
	-\hat{k}^i\sigma^0_{\alpha\dot{\beta}}\bar{\sigma}^{i\dot{\beta}\gamma}h^s_\gamma(\hat{k})=sh^s_\alpha(\hat{k})~,\qquad h^{s\dagger}_{\dot{\gamma}}(\hat{k})\bar{\sigma}^{0\dot{\gamma}\beta}\sigma^i_{\beta\dot{\alpha}}\hat{k}^i=sh^{s\dagger}_{\dot{\alpha}}(\hat{k})~.
\end{align}
The orthonormal condition is
\begin{align}
	h^{s\dagger}_{\dot{\alpha}}\bar{\sigma}^{0\dot{\alpha}\beta}h^{s'}_\beta=\delta^{ss'}~,\qquad \sum_sh^s_\alpha h^{s\dagger}_{\dot{\beta}}\bar{\sigma}^{0\beta\dot{\beta}}=\delta^\beta_\alpha~.
\end{align}
The (anti)-commutation relations are
\begin{align}
	\{b_{\mathbf{k}}^s,b^{s'\dagger}_{\mathbf{k'}}\}=\{d_{\mathbf{k}}^s,d^{s'\dagger}_{\mathbf{k'}}\}=(2\pi)^3\delta^{ss'}\delta^3(\mathbf{k}-\mathbf{k'})~,
\end{align}
while all other anticommutators vanish. The mode functions can be solved exactly in the form of the Whittaker $W$-function\footnote{Notice that two of our mode functions differ from those of \cite{Hook:2019zxa} by a sign. This is because our ``Dirac mass'' is $-\Delta$ in (\ref{SkappaDelta}). This difference is purely a matter of convention, and physical results do not depend on it.} 
\begin{align}\label{modefunction}
	\nonumber&u_+=\frac{H^{3/2}\tilde{\Delta}e^{-\pi\tilde{\kappa}/2}}{\sqrt{2k}}\tau\, W_{-\frac{1}{2}+i\tilde{\kappa},i\tilde{\mu}}(2ik\tau)~,\qquad	u_-=-\frac{iH^{3/2}e^{\pi\tilde{\kappa}/2}}{\sqrt{2k}}\tau\, W_{\frac{1}{2}-i\tilde{\kappa},i\tilde{\mu}}(2ik\tau)~,\\
	&v_+=-\frac{iH^{3/2}e^{-\pi\tilde{\kappa}/2}}{\sqrt{2k}}\tau\, W_{\frac{1}{2}+i\tilde{\kappa},i\tilde{\mu}}(2ik\tau)~,\qquad~~~~	v_-=\frac{H^{3/2}\tilde{\Delta}e^{\pi\tilde{\kappa}/2}}{\sqrt{2k}}\tau\, W_{-\frac{1}{2}-i\tilde{\kappa},i\tilde{\mu}}(2ik\tau)~,
\end{align}
where we have denoted $\tilde{\Delta}\equiv\Delta/H,~ \tilde{\kappa}\equiv\kappa/H,~ \tilde{\mu}\equiv\mu/H$ and $\mu\equiv\sqrt{\Delta^2+\kappa^2}$. By either the IR expansion \cite{Adshead:2015kza} or the Stokes-line method \cite{Sou:2021juh}, one can find the occupation number for these fermions to be
\begin{align}
	\langle n_{\pm}(\mathbf{k})\rangle'=\langle \bar{n}_{\pm}(\mathbf{k})\rangle'= e^{-2\pi(\tilde{\mu}\pm\tilde{\kappa})}~,\label{dSfermionOccupationNumber}
\end{align}

Now we are in a position to compute the effective potential. To reiterate, we wish to compute
\begin{align}
	S_{\text{eff}}[\Delta]=-\int d^4 x\sqrt{-g}\frac{M^2\Delta^2}{4}-i \ln Z_f[\Delta]~,
\end{align}
In the uniform limit, the effective action is simply a potential function integrated over spacetime volume:
\begin{align}
	S_{\text{eff}}[\Delta]&\approx-\int d^4x\sqrt{-g}V_{\rm{eff}}(\Delta)~,	
\end{align}
with 
\begin{align}
	V_{\rm{eff}}(\Delta)=\frac{1}{4}M^2\Delta^2+V_{\rm{1-loop}}(\Delta)~,\quad\text{and}\quad V_{\rm{1-loop}}(\Delta)=\frac{i\ln Z_f[\Delta]}{\int d^4x\sqrt{-g}}~.\label{singleCouplingVeff}
\end{align}
Now take a derivative with respect to the gap field $\Delta$, we obtain
\begin{align}
	\frac{\partial V_{\mathrm{1-loop}}}{\partial \Delta} &=\frac{1}{\int d^4x\sqrt{-g}}\frac{i\partial\ln Z_f}{\partial\Delta}=-\langle\bar{\psi}\psi\rangle_f~,
\end{align}
where $\langle\cdots\rangle_{f}$ stands for the average value computed using the theory action $S_{f}[\psi]$ in (\ref{SkappaDelta}). Switching to the 2-component spinor variables, the potential derivative becomes
\begin{align}
	\frac{\partial V_{\mathrm{1-loop}}}{\partial \Delta}=-\langle\bar{\psi}\psi\rangle_{f}=-\langle\chi\zeta+\zeta^\dagger\chi^\dagger\rangle_f=\int\frac{d^3k}{(2\pi)^3}\sum_{s=\pm}\bigg[u_s(\tau,k)v^*_s(\tau,k)+u^*_s(\tau,k)v_s(\tau,k)\bigg]~.
\end{align}
We can change the loop momentum variable into a dimensionless variable that corresponds to the physical momentum measured in Hubble units,
\begin{align}
	z\equiv-k\tau=\frac{k/a(\tau)}{H}~.
\end{align}
The potential derivative now takes a more compact form,
\begin{align}
	\frac{\partial V_{\mathrm{1-loop}}}{\partial \Delta}=-\frac{\tilde{\Delta}H^3}{4\pi^2}\left[e^{-\pi\tilde{\kappa}}\mathcal{I}_++e^{\pi\tilde{\kappa}}\mathcal{I}_-+\mathrm{c.c.}\right]~,
\end{align}
where the key integrals are given by
\begin{align}\label{keyint}
	\mathcal{I}_+&\equiv -i\int^{\infty}_0dz\,z W_{-\frac{1}{2}+i\tilde{\kappa},i\tilde{\mu}}(-2iz)\,W_{\frac{1}{2}-i\tilde{\kappa},-i\tilde{\mu}}(2iz)~,\\
	\mathcal{I}_-&\equiv i\int^{\infty}_0dz\,z W_{\frac{1}{2}-i\tilde{\kappa},i\tilde{\mu}}(-2iz)\,W_{-\frac{1}{2}+i\tilde{\kappa},-i\tilde{\mu}}(2iz)~.
\end{align}
In the high-energy limit with $z\rightarrow\infty$, we have $W_{-\frac{1}{2}\pm i\tilde{\kappa},i\tilde{\mu}}\,W_{\frac{1}{2}\mp i\tilde{\kappa},-i\tilde{\mu}}\sim\mathcal{O}(1)$, suggesting these integrals (and the potential derivative itself) are quadratically divergent, as one should expect from the flat spacetime loop integral. As in the flat spacetime case, we can keep the explicit cutoff but understand it as an alternative way of expressing the renormalized quantities. Thus the divergent part of the loop integral is trivial in this sense. As we shall see later, the non-trivial aspect of the dS loop integral is hidden in the \textit{finite} part, encoding the information about the non-perturbative spacetime curvature effects that is not visible at any orders in the $H^n$-expansion. Using a regularization trick based on the Whittaker product integral of Arthur Erdelyi~\cite{Gradshteyn:1702455}, we are able to compute the finite remainder of the loop integral \textit{exactly} for \textit{all} parameter sizes. We spell out the detailed computation of the loop integrals $\mathcal{I}_\pm$ in Appendix~\ref{dSloopApp}, and quote the final result of the effective potential here:
\begin{keyeqn}
	\begin{align}
		\nonumber V_{\rm{eff}}(\Delta)
		&=\frac{1}{4}\left(M^2-\frac{\Lambda^2}{\pi^2}\right)\Delta^2-\left(\frac{\kappa^2-H^2/2}{2\pi^2}\Delta^2-\frac{\Delta^4}{4\pi^2}\right)\ln\frac{\Lambda}{H}\\
		&\quad -\int d\Delta \frac{\Delta H^2}{2\pi^2}\Re\left[f^{\epsilon}_+ +f^{\epsilon}_- +2A+\frac{2B\Delta^2}{H^2} \right]~.\label{VeffdSexact}
	\end{align}
\end{keyeqn}
The analytical expression for the finite remainder $f_{\pm}^\epsilon$ and the coefficient functions $A,B$ are given by (\ref{finiteRemainder}) and (\ref{ABcoefficientFunctions}) in Appendix~\ref{dSloopApp}. Notice that the only divergent dS correction is the logarithmic term $\sim H^2\Delta^2\ln\Lambda$ in the first line, all other corrections come from the finite remainder in the second line.

It is useful to take an $H\to 0$ limit of $V_{\rm{eff}}$ to examine the corrections of slow spacetime expansion. Note that the finite part of the effective potential contains terms non-perturbative in the small-$H$ expansion, as seen from the trigonometric functions in $f_\pm^\epsilon$. To extract the leading non-perturbative correction, we expand the trigonometric functions into exponential functions and only keep terms with the leading exponent. For example, with $\tilde{\mu}=\sqrt{\Delta^2+\kappa^2}/H\to\infty$,
\begin{align}
	\csch(2\pi\tilde{\mu})=\frac{2}{e^{2\pi\tilde{\mu}}-e^{-2\pi\tilde{\mu}}}=2 e^{-2\pi\tilde{\mu}}+\cdots~.
\end{align}
Expanding to the leading non-trivial order in this way, we obtain
\begin{keyeqn}
	\begin{align}
		\nonumber V_{\rm{eff}}(\Delta)=&\,\frac{1}{4}\left[M^2-\frac{\Lambda^2}{\pi^2}+\frac{2\kappa^2(1-\ln 2)}{\pi^2}\right]\Delta^2+\frac{\kappa^2}{2\pi^2}\Delta^2\ln\frac{\Delta}{\Lambda}-\frac{\Delta^4}{8\pi^2}\left(\ln\frac{\Delta}{\Lambda}+\frac 14-\ln 2\right)\\
		\nonumber&-\frac{H^2}{4\pi^2}\left[\left(\frac{25}{48}+\gamma\right)\Delta^2+\Delta^2\ln\frac{\Delta}{\Lambda}-3\kappa^2\ln\frac{\Delta}{\kappa}+\frac{\kappa^4}{3\Delta^2}\right]\\
		\nonumber&+\mathcal{O}(H^4)\\
		\nonumber&+e^{-\frac{\sqrt{\Delta^2+\kappa^2}-\kappa}{H/(2\pi)}}\frac{H}{2\pi}\Bigg\{\frac{\kappa ^3}{2 \pi ^2}\left[3 \left(1+\frac{\Delta ^2}{\kappa ^2}\right)-\left(2-\frac{\Delta ^2}{\kappa ^2}\right) \sqrt{1+\frac{\Delta ^2}{\kappa ^2}} \mathrm{arccoth}
		\sqrt{1+\frac{\Delta^2}{\kappa^2}}\,\right]\\
		\nonumber&\qquad\qquad\qquad\qquad +\mathcal{O}(H)\Bigg\}\\
		&+\mathcal{O}\left(e^{-2\pi(\tilde{\mu}+\tilde{\kappa})}\right)~.\label{VeffdSExpanded}
	\end{align}
\end{keyeqn}	
Here the first line reproduces the flat-spacetime result in (\ref{flatV}) after taking $H\to 0$. The second and third lines give perturbative corrections \textit{analytic} in $H$, while the fourth and fifth lines correspond to the non-perturbative corrections that are \textit{non-analytic} in $H$. Inspecting the form of the $\mathcal{O}(H^2)$ corrections, we find that there is a superficial \textit{dip} dominating the small-gap region where $\Delta^2\ll H \kappa$. Naively, this dip seems to suggest that the effective potential is unbounded from below in the gapless regime, and that no stable BCS condensation can occur for an arbitrarily small expansion rate. However, one must recall that the expansion (\ref{VeffdSExpanded}) is organized in powers of $H$, and it would be inconsistent to take $\Delta^2\ll H \kappa$ without considering all the higher order contributions. In fact, a full inspection using the exact effective potential (\ref{VeffdSexact}) shows the effect of such terms is actually to lower the potential barrier at the origin. Therefore, raising the Hubble parameter leads to the formation of a true vacuum at $\Delta=0$, and eventually to a symmetry-restoring phase transition and the disintegration of BCS condensation. We will come back to the details of BCS phase transition in Sect.~\ref{phaseTransitionSect}.

Note that with the benefit of hindsight, we have dressed the non-analytic Hubble corrections in (\ref{VeffdSExpanded}) in the form of the dS background temperature
\begin{align}
	T_{\mathrm{dS}}=\frac{H}{2\pi}~.
\end{align}
Thus the non-analytic pieces are essentially suppressed by the corresponding Boltzmann factors. Take the first non-analytic piece for example, its Boltzmann factor reads
	\begin{align}
	e^{-2\pi(\tilde{\mu}-\tilde{\kappa})}=e^{-\frac{\sqrt{\Delta^2+\kappa^2}-\kappa}{H/(2\pi)}}\sim \frac{1}{e^{(\mu-\kappa)/T_{\mathrm{dS}}}+1}~,
\end{align}
which is the occupation number of the fermionic excitation upon the BCS ground state (more precisely, that of the enhanced left-handed fermions/anti-fermions as in (\ref{dSfermionOccupationNumber})). Here $\mu= \sqrt{\Delta^2+\kappa^2}$ is the effective mass of the fermions. From the analogy of BCS theory at a finite temperature, we recognize this non-analytic piece as the on-shell particle production effect of the dS background temperature. Of course, one should not take the thermal analogy too seriously, because there are power-law terms such as $H^{n}\sim T_{\mathrm{dS}}^{n}$ in (\ref{VeffdSExpanded}), which should be absent in a gapped thermal system at low temperatures. These are more naturally viewed as generic spacetime curvature corrections that are not necessarily dS-like.

\paragraph{Backreaction on inflation} At the end of this subsection, we point out that although non-perturbative information is encoded in (\ref{VeffdSexact}), we are nevertheless working in the mean-field approximation (whose validity will be checked later in Sect.~\ref{CollectiveModeSubSect}) and have so far neglected the backreaction on the inflaton and graviton sector. Thus (\ref{VeffdSexact}) is exact in a restricted sense. To estimate how much the BCS matter sector backreacts to the inflationary background, we can compare the typical energy deposit in the fermion condensation and the vacuum energy stored in the inflaton background. The former is estimated from the change of effective potential (\ref{VeffdSExpanded}) during a typical gap field traverse $\delta \ln\Delta=\mathcal{O}(1)$,
\begin{align}
	\delta V_{\text{eff}}\approx \frac{\kappa^2 \Delta_0^2}{2\pi^2}~,
\end{align}
where we have chosen the second term in the expanded effective potential (\ref{VeffdSExpanded}) as a benchmark of its dependence on the gap. In slow-roll inflation, the vacuum energy in the inflaton background is given in terms of the Hubble parameter and the Planck scale as
\begin{align}
	\rho_{\text{inf}}\approx 3M_p^2 H^2\approx\frac{3}{2}\frac{\dot{\phi}_0^2}{\varepsilon}~,
\end{align}
where $\varepsilon\equiv -\dot{H}/H^2\approx \dot{\phi}_0^2/(2M_p^2 H^2)\ll 1$ is the first slow-roll parameter. Therefore, we have
\begin{align}
	\frac{\delta V_{\text{eff}}}{\rho_{\text{inf}}}\approx\frac{1}{3\pi^2}\times\varepsilon\times\frac{\kappa^2\Delta_0^2}{\dot{\phi}_0^2}\lesssim\frac{1}{12\pi^2}\times \varepsilon\times\frac{\kappa^4}{\dot{\phi}_0^2}\lesssim\frac{\varepsilon}{12\pi^2}\ll 1~.\label{LatentHeatFractionEstimate}
\end{align}
Here in the second step we have used $\Delta_0\lesssim \kappa/2$, which is found to be satisfied in most of the parameter region of the condensed phase. The last step comes from requiring $\kappa^2<\dot{\phi}_0<\Lambda_c^2$, which is itself a consequence of perturbative unitarity \cite{Chen:2018xck,Wang:2019gbi}. Henceforth we see that the phase transition in the fermion sector is negligible compared to the background inflaton dynamics, consistent with neglecting backreaction.


\subsection{Other self-interactions}\label{U1A}	

We have so far focused on the four-fermion interaction $(\bar\psi\psi)^2$ in the $s$-channel pairing. In principle, one can also consider other types of fermion self-interactions.
At the level of dimension-6 operators, the most general set of Lorentz-invariant and $CPT$-invariant operators that couple fermion and anti-fermions is given by squaring the five basic bilinears\footnote{The Lorentz structure dictates the contraction of Lorentz indices, while the mixing between scalar (vector) and pseudoscalar (pseudovector) is forbidden by $P$.}:
\begin{align}
\left\{(\bar{\psi}\psi)^2~,\quad(i\bar{\psi}\gamma^5\psi)^2~,\quad(\bar{\psi}\gamma^\mu\psi)^2~,\quad(\bar{\psi}\gamma^\mu\gamma^5\psi)^2~,\quad(\bar{\psi}\sigma^{\mu\nu}\psi)^2\right\}~.
\end{align}
However, due to Fierz identities, only three of the five combinations are linearly independent \cite{1937ZPhy,Bertrand2005ARB}. We are free to choose the scalar, pseudoscalar and vector combinations to span the space of quartic interactions. We need to keep the spatial isotropy in any BCS condensations, so the vector condensate will be excluded\footnote{In principle, one can also introduce a vector BCS condensate while preserving isotropy, by including an $SU(2)_I$ internal symmetry~\cite{Endlich:2012pz,Nicolis:2015sra}. Then the expectation value of the order parameter breaks the $SU(2)_I\times SO(3)$ to the diagonal subgroup of effective rotations, which is still a symmetry. We leave this interesting scenario to future studies.}, leaving us with only the scalar and pseudoscalar combinations. Note that in the inflationary context, we do not necessarily impose so many symmetries. The rolling speed of the inflaton spontaneously breaks the boosts (potentially also $P$ and $T$) and introduces a distinction between the temporal and spatial components of fields. However, we do not wish to introduce such a distinction at the level of dimension-6 operators, because this would reintroduce the explicit inflaton derivative, i.e. $(\bar{\psi}\gamma^0\psi)^2\subset (\partial_\mu\phi\bar{\psi}\gamma^\mu\psi)^2$, which dramatically increases the mass dimension of the operator.

Henceforth, we will simply generalize our model to include the pseudoscalar self-coupling,
\begin{align}
\mathcal{L}_{\rm{int}}'=\frac{(\bar{\psi}\psi)^2}{M^2}+\frac{(\bar{\psi}i\gamma^5\psi)^2}{M_5^2}~.\label{dS2couplingModel}
\end{align}
In the equal-coupling case with $M=M_5$, the theory enjoys a chiral $U(1)_A$ symmetry:
\begin{align}
\psi&\rightarrow e^{i\alpha \gamma^5}\psi,\\
\bar{\psi}&\rightarrow\bar{\psi}e^{i\alpha\gamma^5}~.
\end{align}
This $U(1)_A$ will be spontaneously broken whenever the fermions form a BCS condensate.	Now, we need to introduce two auxiliary scalar fields to resolve the self-interactions
\begin{align}
Z=\int \mathcal{D}\psi\mathcal{D}\bar\psi\, e^{i S_{\text{dS}}'[\psi]}=\int \mathcal{D}\Sigma\mathcal{D}\Pi e^{i S_{\text{eff}}[\Sigma,\Pi]}~,
\end{align}
with
\begin{align}
S_{\text{eff}}[\Sigma,\Pi]=-\int d^4 x \sqrt{-g} \frac{M^2\Sigma^2+M_5^2\Pi^2}{4}-i \ln Z_f[\Sigma,\Pi]~,
\end{align}
and
\begin{align}
Z_f[\Sigma,\Pi]=\int\mathcal{D}\psi\mathcal{D}\bar\psi\, e^{i S'_f[\psi]}~,~
S_f'[\psi]=\int d^4x\sqrt{-g}~\bar{\psi}\left(i\gamma^a e^\mu_{~~a}\nabla_\mu-\kappa\gamma^0\gamma^5+\Sigma+i\gamma^5\Pi\right)\psi~.\label{Sfprime}
\end{align}
In the uniform limit, $\Sigma(x),\Pi(x)=\mathrm{const}$, we can always perform a $U(1)_A$ rotation $\psi\rightarrow e^{-i\Theta \gamma^5/2}\psi$ to rotate away the mass phase,
\begin{align}
\bar{\psi}\left(\Sigma+i\gamma^5\Pi\right)\psi\rightarrow\bar{\psi}e^{-i\Theta \gamma^5}\left(\Sigma+i\gamma^5\Pi\right)\psi=\Delta\bar{\psi}\psi~.
\end{align}
In doing so, one naturally performs a field redefinition
\begin{align}
\Delta&\equiv\sqrt{\Sigma^2+\Pi^2}~,\\
\Theta&\equiv\arctan \frac{\Pi}{\Sigma}~.
\end{align}
Now the fermion partition function $Z_f$ takes exactly the same form of (\ref{SkappaDelta}). The effective IR theory is thus turned into
\begin{align}
Z=\int \mathcal{D}\Delta\mathcal{D}\Theta\, \mathcal{J} e^{i S_{\text{eff}}[\Delta,\Theta]}=\int \mathcal{D}\Delta\mathcal{D}\Theta \mathcal{D}c \mathcal{D}\bar{c} \,e^{i S_{\text{eff}}[\Delta,\Theta]+i\int d^4 x \Delta \bar{c}c}~,\label{Seffgamma5}
\end{align}
where
\begin{align}
S_{\text{eff}}[\Delta,\Theta]=-\int d^4 x \sqrt{-g} \frac{M^2\cos^2\Theta+M_5^2\sin^2\Theta}{4}\Delta^2-i \ln Z_f[\Delta]~.
\end{align}
We have also introduced the auxiliary ghost fields $c,\bar{c}$ to replace the Jacobian $\mathcal{J}\equiv\det[\partial(\Sigma,\Pi)/\partial(\Delta,\Theta)]$.

The ground state solution is obtained at the stationary points of the path integral (\ref{Seffgamma5}). For the non-dynamical ghost fields, this is simply attained at $c=\bar{c}=0$. A subsequent variation with respect to the uniform gap fields $\Delta,\Theta$ then selects the BCS vacuum, according to the minima of the effective potential
\begin{align}
V_{\text{eff}}(\Delta,\Theta)=\frac{M^2\cos^2\Theta+M_5^2\sin^2\Theta}{4}\Delta^2+V_{\mathrm{1-loop}}(\Delta)~,
\end{align}
where $V_{\mathrm{1-loop}}(\Delta)$ is identical to the case computed in the previous section. Inspecting the dependence of the effective potential on the coupling strengths, we find three limiting scenarios:
\begin{enumerate}
	\item $M=M_5$. In this equal-coupling case, the effective potential reduces to that of (\ref{singleCouplingVeff}) in the single-coupling case, which is a function of the amplitude field $\Delta$ only. The phase field $\Theta$ now enjoys a shift symmetry $\Theta\to \Theta+C$, a legacy of the UV chiral $U(1)_A$. Upon acquiring a non-zero average value with $\Delta_0\neq 0$, the phase mode $\Theta$ becomes the massless Goldstone boson of the spontaneous broken shift symmetry.
	\item $|M-M_5|\ll M$. As one deviates from the equal-coupling limit, the phase mode acquires a soft mass proportional to the coupling difference: $m_{\Theta}^2\propto M^2-M_5^2$.
	\item $M\ll M_5$ (or $M\gg M_5$). When there is a hierarchy between the couplings, the gap fields automatically align along the maximal symmetry breaking direction. In other words, the phase field becomes frozen at $\Theta=0$ ($\Theta=\pi/2$) and decouples from the spectrum due to its heavy mass. Thus the theory reduces to the single-coupling model in the previous section.
\end{enumerate}


\section{BCS phase transition during inflation}\label{phaseTransitionSect}

Equipped with the exact formula of the effective potential, let us now analyze the parameter dependence of this effective potential and the phase diagram of the inflationary BCS condensation. Without loss of generality, we will focus on the single-coupling model (\ref{dSMainModel}), and the result can be directly translated to the more general two-coupling model (\ref{dS2couplingModel}).

The effective potential $V_{\text{eff}}(\Delta)$ is dependent on four parameters: the chemical potential $\kappa$, the Hubble scale of inflation $H$, the coupling strength $M$, and the cutoff scale $\Lambda$ which is understood as playing the role of mass renormalization parameter. Without referring to any external energy scales such as the Planck scale $M_p\sim 10^{18}$GeV, the property of the BCS system only depends on the \textit{relative} size of these four parameters. We need to determine which of them should be fixed and which of them are to be considered tunable.

In a realistic inflationary spacetime, the background has a weak time dependence characterized by the slow roll parameters $\varepsilon,\eta$, introduce a weak time dependence in the chemical potential and the Hubble parameter,
\begin{align}
	\kappa(t)=\frac{\dot{\phi}_0(t)}{\Lambda_c}\approx\frac{\dot{\phi}_0(t_*)}{\Lambda_c}\left(1+\frac{\eta}{2} H_*(t-t_*)\right)~,\quad H(t)\approx H_{*}\left(1-\varepsilon H_*(t-t_*)\right)~,
\end{align}
where $t_*$ is a reference time during inflation. In the slow-roll phase of inflation, $\varepsilon,\eta>0$, thus the chemical potential tends to increase, while the Hubble parameter tends to decrease during inflation. With this observation in mind, we will mainly study our BCS model dependence on the chemical potential and Hubble parameter, while holding the coupling and cutoff fixed. In all cases, the Hubble scale turns out to be the smallest scale of the problem, thus we will typically quantify other scales in units of Hubble. When we need to vary the Hubble scale, we will use a reference Hubble scale $H_*$ to quantify other dimensional parameters.

Before a quantitative discussion, recall that in the flat spacetime case, the gap equation always admits a finite but exponentially small solution. Therefore, the system is always in the condensed phase. However, with the spacetime expansion effects of inflation, it is not true that the gap equation always admits a non-zero solution. We have seen in the previous section that the dS correction looks thermal at the leading order, with a Hubble temperature $T_{\text{dS}}\sim H$.  For BCS superconductors, when the temperature rises above a critical temperature $T_c$, superconductivity will be lost and the BCS condensation will disintegrate
\begin{align}
	k_B T_c\sim \Delta_{\text{BCS}}~,
\end{align}
Borrowing this analogy, during inflation, this should happen at
\begin{align}
	k_B T_{\text{dS}}\sim \Delta_0~,
\end{align}
and the explanation here is that spacetime expansion tears the Cooper pairs apart, destroying the condensate. This estimation gives a loose lower bound on the gap size, $\Delta_0\gtrsim H/(2\pi)$. We will see later the actual lower bound is tighter.

\begin{figure}[htp] 
	\centering
	\includegraphics[width=15cm]{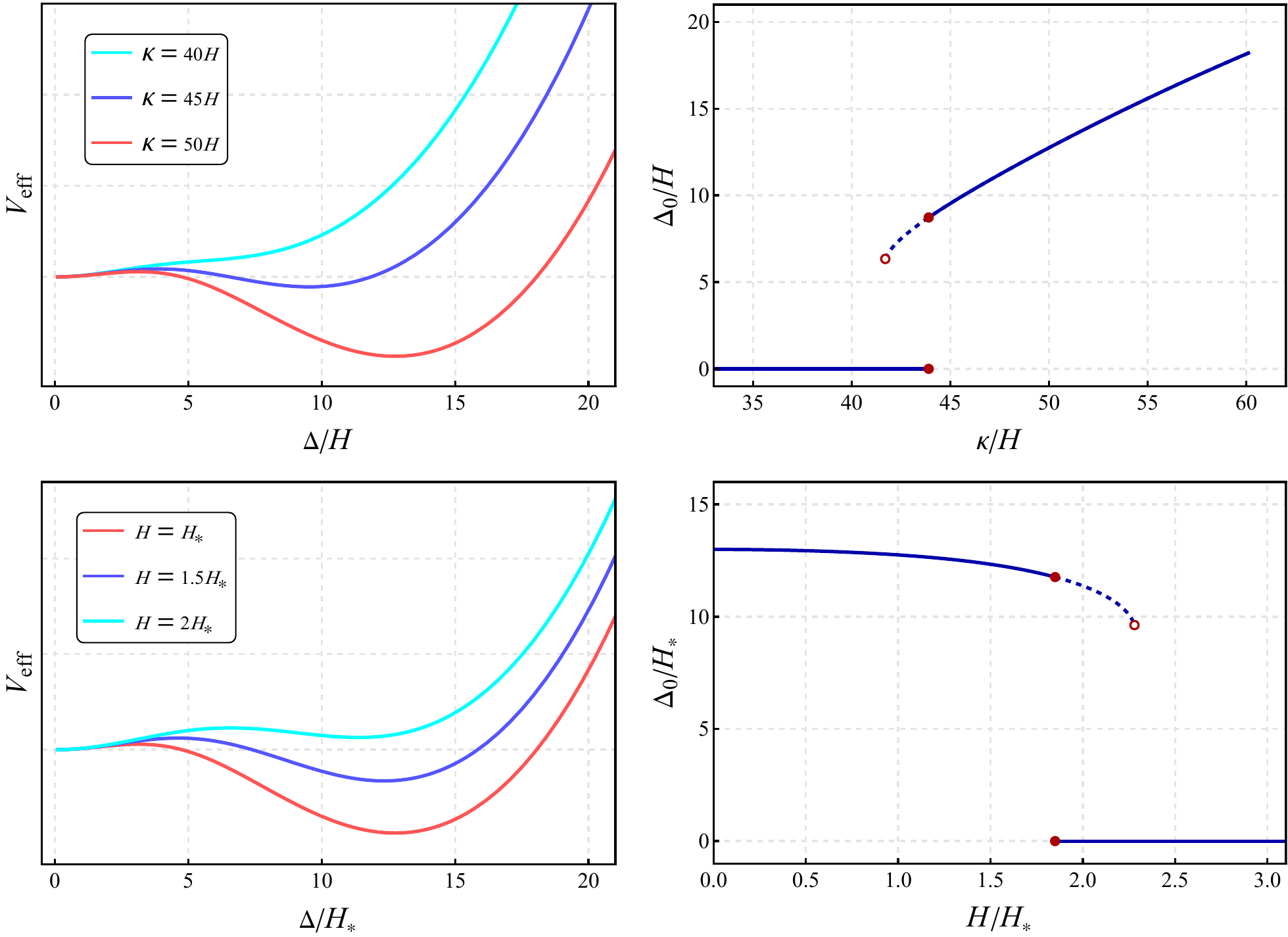}
	\caption{\label{veffplot} Left column: The effective potential for different choices of chemical potential (first row) and Hubble parameter (second row), with the other parameters fixed to be $M=40H,\Lambda=100H$. Right column: The BCS gap for different choices of chemical potential (first row) and Hubble parameter (second row), with the other parameters fixed to be $M=40H_*,\Lambda=100H_*,\kappa=50H_*$. In the condensed phase, the BCS gap increases with chemical potential and decreases with the Hubble parameter. The discontinuous change of the gap signals a first-order phase transition, where the two vacua possess equal free energy but are separated by a potential barrier. Beyond the critical chemical potential/temperature (solid circles), the asymmetric vacuum is metastable and may tunnel back to the symmetric vacuum via bubble nucleation that dismantles the condensate. The empty circles correspond to the inflection point where the false vacuum disappears.}
\end{figure}

\begin{figure}[htp] 
	\centering
	\includegraphics[width=10cm]{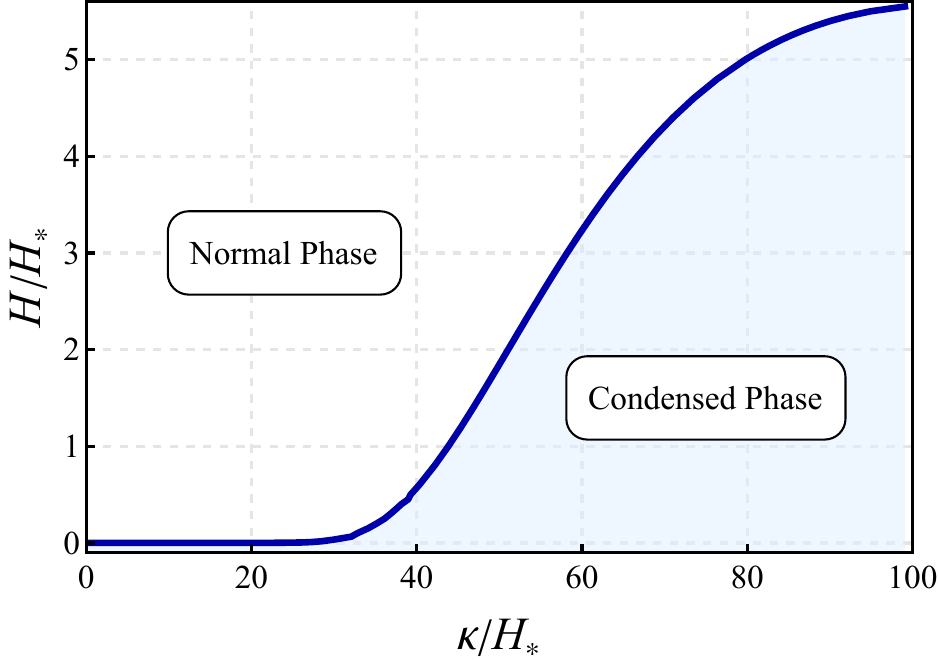}
	\caption{\label{Phasedigram} The phase diagram of our inflationary BCS model. The solid blue line separates the normal phase and the condensed phase, all the way from the origin to the cutoff scale $\Lambda=100H_*$. Here the coupling is chosen to be $M=40H_*$. In the exact flat spacetime case with $H=0$, the system always stays in the condensed phase, while an exponentially small Hubble expansion rate can destabilize the condensate.}
\end{figure}

We plot the effective potential and the gap for different sizes of the chemical potential and Hubble parameter in FIG.~\ref{veffplot}. Clearly, at a constant chemical potential, the gap only exists when Hubble is below a critical value. Meanwhile, for a fixed Hubble parameter, the chemical potential has to be large enough to support the existence of the gap. More interestingly, we find that the gap changes in a discontinuous way, and that the phase transition is first order, in agreement with our expectation above. We are also able to draw the phase diagram in FIG.~\ref{Phasedigram}, where the normal phase and condensed phase are discontinuously separated by a curve that extends all the way from the origin to the cutoff.

In fact, if we assume several reasonable conditions
\begin{align}
	\nonumber\text{Naturalness:}&\quad M^2-\frac{\Lambda^2}{\pi^2}>\frac{M^2}{4}~,\\
	\text{EFT validity:}&\quad 0<\kappa<\Lambda\quad\text{and}\quad M,\Lambda>H~,\label{NaturalnessAndEFTValidityConds}
\end{align}
we can give a numerical proof that the phase transition is always first order. Here naturalness means the loop correction to the mass operator $M^2 \Delta^2$ is not much larger than the tree-level value\footnote{As mentioned in Sect.~\ref{flatSpacetimeSect}, $\Lambda$ is understood as a physical parameter determined through implicit renormalization conditions. Therefore it is not meaningful to directly compare $\Lambda$ to $M$ and conclude about EFT validity. Instead, one should look at the loop correction of the potential in the large-$\Delta_0$ regime and require it to be less than the tree-level contribution. Therefore, EFT validity is still preserved as along as $\Lambda<\pi M$. This leaves room for $\Lambda$ being greater than $M$ without breaking EFT validity. On the other hand, if one goes beyond $\Lambda=\pi M$, the effective potential is overturned and a tachyonic instability appears signalling the breakdown of EFT. To avoid fine tuning, we have also imposed a naturalness bound here to avoid being dangerously close to instability.}, typically not by a factor of order two for $M$. Also, the fermion self-interactions must be established as an EFT operator above the Hubble scale. The proof is simple: second-order phase transitions happen when the second-order derivative of the effective potential vanishes,
\begin{align}
	\frac{\partial^2 V_{\text{eff}}}{\partial\Delta^2}\Bigg|_{\Delta=0}=0~.\label{2ndOrderPTCondition}
\end{align}
However, it is a straightforward although technical task to show that the above equation has no solution under the set of reasonable conditions (\ref{NaturalnessAndEFTValidityConds}). More explicitly, we can numerically solve the minimum of the second-order derivative of the effective potential,
\begin{align}
	\nonumber\min_{\kappa,M,\Lambda,H\in (\ref{NaturalnessAndEFTValidityConds})} \frac{\partial^2 V_{\text{eff}}}{\partial\Delta^2}\Bigg|_{\Delta=0}
	=&\min_{\kappa,M,\Lambda,H\in (\ref{NaturalnessAndEFTValidityConds})} \frac{1}{2}\Bigg\{M^2-\frac{\Lambda^2+2\kappa^2\ln 2}{\pi^2}+\frac{\left(2 \kappa ^2-H^2\right)}{\pi^2}\Re\left[\psi\left(\frac{2 i \kappa }{H}\right)-\ln \frac{\Lambda }{H}\right]\\
	\nonumber&\qquad\qquad\qquad+\frac{3 H \kappa  \coth (2 \pi  \kappa/H )}{2 \pi }-\frac{(27+16 \gamma ) H^2}{16 \pi ^2}\Bigg\}\\
	=&\quad 0.128...\times H^4>0~.
\end{align}
Therefore, $V''_{\text{eff}}(0)=0$ has no solution in the domain (\ref{NaturalnessAndEFTValidityConds}) defined by naturalness and EFT validity, and second-order phase transition is impossible. Note that the above proof is a sufficient condition for first-order phase transition, but is sensitive to the naturalness bound tightness. It will be interesting to investigate the phase transition type more systematically under even more relaxed constraints in future works.

At last, we briefly comment on the two-coupling model case. As mentioned in the previous section, the order parameter $\Theta$ automatically picks the smaller of $\{M^2,M_5^2\}$ to align along. All the discussions above are then directly translated via the replacement $M^2\to\min\{M^2,M_5^2\}$.
	

\section{Observational signatures} \label{ObsSect}

Our BCS model in inflation also brings characteristic observational signatures in the sky (hence the title). First, in the condensed phase, the excitation modes of the BCS vacuum couple to the inflaton through the chemical potential operator, and lead to cosmological collider signals that probe the particle spectrum of the system. Such oscillatory signals can be observed in the squeezed limit of the primordial curvature bispectrum. Second, if the universe starts out in the normal phase, due to the slow time dependence of the inflaton rolling speed, the increasing chemical potential tends to drive the fermions to the condensed phase via a first-order phase transition. Such a phase transition is accompanied by bubble nucleations and collisions, producing stochastic gravitational waves. In this section, we will directly adopt the two-coupling model and investigate these two potential observables.

\subsection{Collective modes}\label{CollectiveModeSubSect}

Up till now, we have been focusing on the background solution of the BCS model, namely, that of the vacuum selection. The state in reality is, however, not the exact BCS vacuum. One can view this as due to the on-shell particle production in an expanding spacetime. We have seen that the non-perturbative correction to the effective potential does include Boltzmann-suppressed real-particle effects. The collective modes describing fluctuations of the Cooper-pair condensate in magnitude and in phase are known as the amplitude mode (or Higgs mode) and phase mode (or Anderson-Bogoliubov mode), respectively~\cite{Pekker, anderson,Bogolyubov:1958se,Anglani:2011cw}.

To analyze these collective modes, we need to perturb the state around the BCS vacuum, and look at the effective action at next-to-leading order in the gradient expansion,
\begin{align}
	S_{\text{eff}}[\Delta,\Theta]=\int d^4x \sqrt{-g}\left[-V_{\text{eff}}(\Delta,\Theta)+\mathcal{Z}_{\Delta}^{\mu\nu}\partial_\mu\Delta\partial_\nu\Delta +\mathcal{Z}_{\Theta}^{\mu\nu}\partial_\mu\Theta\partial_\nu\Theta+\mathcal{O}(\partial^4)\right]~.
\end{align}
The free dynamics of these modes will then be dictated by the quadratic Lagrangian expanded from the above effective action. Since here we are interested in probing the perturbation dynamics, we will assume a hierarchy between the gap and the Hubble scale, $\Delta_0\gg H$ such that the flat spacetime approximation is reliable. This will significantly simplify our calculation. Due to the boost-breaking chemical potential, we do not expect $\mathcal{Z}_{\Delta}^{\mu\nu},\mathcal{Z}_{\Theta}^{\mu\nu}\propto g^{\mu\nu}$, and these scalar modes will acquire a non-unit sound speed in general.

In the flat spacetime limit $\mathcal{O}(H^0)$, the field strength renormalization coefficients are easily solved by computing 1-loop diagrams,
\begin{align}\label{Zcoeff}
	\mathcal{Z}_{\sf{a}}^{00}=-i\frac{d}{dk_0^2}\left(\parbox{0.2\textwidth}{\includegraphics[width=0.2\textwidth]{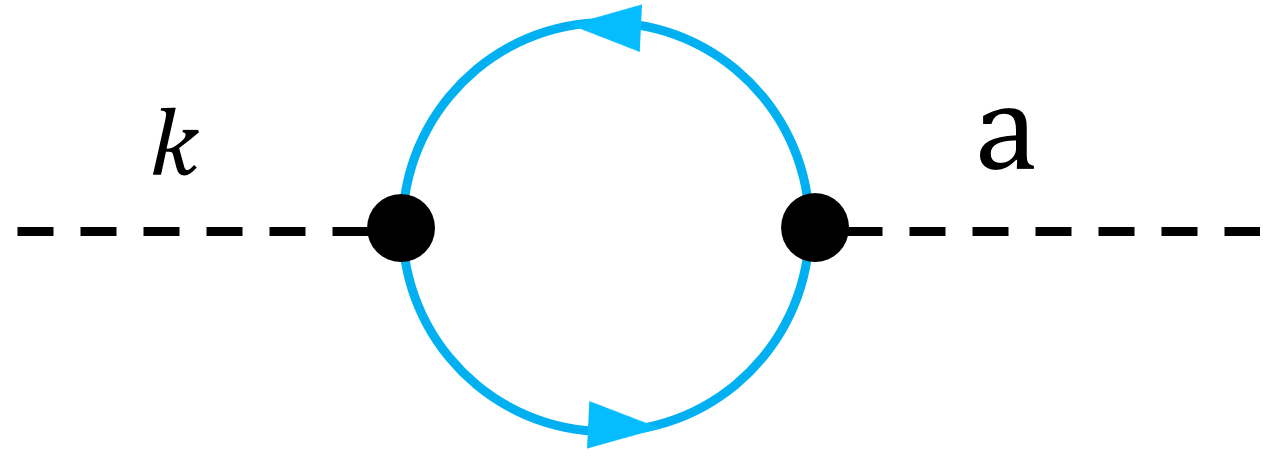}}\right)\Bigg|_{k_0\rightarrow 0},~
	\mathcal{Z}_{\sf{a}}^{33}=-i\frac{d}{d|\mathbf{k}|^2}\left(\parbox{0.2\textwidth}{\includegraphics[width=0.2\textwidth]{Zsigmaortheta}}\right)\Bigg|_{|\mathbf{k}|\rightarrow 0}~,
\end{align}
Here the subscript stands for $\sf{a}=$ $\Delta,\Theta$. Regulating the logarithmically divergent integrals using a hard cutoff, we find (see more details in Appendix~\ref{ZloopApp})
\begin{align}
	\mathcal{Z}_\Delta^{00}=&\frac{\kappa^2}{12\pi^2\Delta^2}\left(1-\frac{3\Delta^2}{\kappa^2}\ln \frac{\Delta}{\Lambda_\sigma}\right)+\mathcal{O}(H^2)~,\label{Z00formula}\\
	\mathcal{Z}_\Delta^{33}=&-\frac{\kappa^2}{12\pi^2\Delta^2}\left[\frac{1}{3}-\frac{3\Delta^2}{\kappa^2}\left(\ln \frac{\Delta}{\Lambda_\sigma}+\frac{1}{3}\right)+\frac{2\Delta^4}{\kappa^3\sqrt{\Delta^2+\kappa^2}}\rm{Arctanh}\frac{\kappa}{\sqrt{\Delta^2+\kappa^2}} \right]+\mathcal{O}(H^2)~,\\
	\mathcal{Z}_\Theta^{00}=&\frac{\kappa^2}{4\pi^2}\left[1-\frac{\Delta^2}{\kappa^2}\ln \frac{\Delta}{\Lambda_{\theta}}\right]+\mathcal{O}(H^2)~,\\ \mathcal{Z}_\Theta^{33}=&-\frac{\kappa^2}{4\pi^2}\left[\frac{1}{3}-\frac{\Delta^2}{\kappa^2}\left(\ln\frac{\Delta}{\Lambda_{\theta}}-\frac{1}{3}\right)\right]+\mathcal{O}(H^2)~.
\end{align}
Inserting this back to the effective action, we perturb the vacuum by the canonically normalized modes,
\begin{align}
	\Delta(x)=\,\Delta_0+\frac{\sigma(x)}{\sqrt{2 \mathcal{Z}_\Delta^{00}(\Delta_0)}}~,\quad
	\Theta(x)=\,\Theta_0+\frac{\theta(x)}{\sqrt{2 \mathcal{Z}_\Theta^{00}(\Delta_0)}}~,
\end{align}
where $\sigma$ and $\theta$ indicate the amplitude and phase modes, respectively. $\Theta_0=0$ ($\Theta_0=\pi/2$) is the vacuum angle when $M\leqslant M_5$ ($M> M_5$). Expanding to quadratic order and restoring the scale factors adiabatically, we obtain
\begin{align}
	\nonumber S_{\text{eff}}=\frac{1}{2}\int d\tau d^3x a^2(\tau)\Bigg[&(\partial_\tau\sigma)^2-c_\sigma^2(\partial_i\sigma)^2-a^2(\tau)m_\sigma^2 \sigma^2\\
	\nonumber+&(\partial_\tau\theta)^2-c_\theta^2(\partial_i\theta)^2-a^2(\tau)m_\theta^2 \theta^2\\
	+&\mathcal{O}(H^2,\partial^4,\sigma^3,\theta^4)\Bigg]~.
\end{align}
This quadratic action is useful for reading out the sound speed and the mass information. Interestingly, we find that in the small gap limit with $\kappa\gg \Delta_0>H$, the sound speeds universally approach to that of a relativistic fluid,
\begin{align}
	c_\sigma^2\equiv-\frac{\mathcal{Z}_\Delta^{33}}{\mathcal{Z}_\Delta^{00}}\Bigg|_{\Delta_0,\Theta_0}\approx\frac{1}{3}~,\qquad c_\theta^2\equiv-\frac{\mathcal{Z}_\Theta^{33}}{\mathcal{Z}_\Theta^{00}}\Bigg|_{\Delta_0,\Theta_0}\approx\frac{1}{3}~.
\end{align}
This is a reminiscence of the sound speed of non-relativistic BCS excitations in superconductors \cite{anderson,Littlewood}. The mass of the collective modes is given by
\begin{align}
	m_\sigma^2\equiv \frac{\partial_\Delta^2 V_{\text{eff}}}{2 \mathcal{Z}_\Delta^{00}}\Bigg|_{\Delta_0,\Theta_0}\approx 6\Delta_0^2~,\qquad
	m_\theta^2\equiv \frac{\partial_\Theta^2 V_{\text{eff}}}{2 \mathcal{Z}_\Theta^{00}}\Bigg|_{\Delta_0,\Theta_0}=\frac{\pi^2 }{\kappa^2}|M^2-M_5^2|\Delta_0^2~.
\end{align}
As mentioned in Sect.~\ref{U1A}, in the equal-coupling case, $m_\theta^2=0$, and the phase mode $\theta$ is the massless Goldstone of $U(1)_A$ symmetry breaking. Otherwise, the $\theta$ mode acquires a mass proportional to the coupling difference. Notice also that the amplitude mode mass $\sqrt{6}\Delta_0$ is slightly larger than that of the non-relativistic case ($2\Delta_0$), although the fermionic excitations still have a gap of $\Delta_0$.

Due to the cosmic expansion, the gap solution cannot be arbitrarily small. We find in most parameter regions, it is of order $\Delta_0\gtrsim\mathcal{O}(10)H$. Thus the amplitude mode is very heavy in Hubble units. This suggests a negligible cosmological collider signal of the amplitude mode. The phase mode can be light in the case of soft $U(1)_A$ breaking, but it turns out that its coupling to inflaton is in general rather weak, hence also a small signal strength, as we will see later.

\paragraph{A check on the mean-field approximation} The calculation of gap equation and the phase analysis above are performed in the mean-field approximation where the gap field is assumed to take a uniform background with small fluctuations. Now that the linear dynamics of the fluctuations $\sigma$ have been understood, we can estimate how well the mean-field approximation works. More specifically, we need to check the Ginzburg criterion for reaching a stable condensed vacuum, 
\begin{align}
	\frac{\langle\sigma^2(x)\rangle}{2\mathcal{Z}_\Delta^{00}}=\langle(\Delta(x)-\Delta_0)^2\rangle<\Delta_0^2~.\label{GinzburgCriterion}
\end{align}
Namely, we require the standard deviation of the quantum fluctuations of the order parameter to be smaller than its mean value. To the leading order in gradient expansion, the two-point function (power spectrum) of $\sigma$ is given by that of a canonical massive scalar field in de Sitter spacetime\footnote{The two-point function of quantum fields is usually singular in the coincident limit and needs renormalization. The leading contributions to $\langle\sigma^2(x)\rangle$ are of order $\mathcal{O}(m^2_\sigma)$ and $\mathcal{O}(H^2)$, but they are renormalization-scheme dependent since they come from UV divergences. The leading contribution that does not depend on renormalization scheme turns out to be of order $\mathcal{O}(H^4/m^2_\sigma)$. It is straightforward to see that even if we evoke the renormalization-scheme dependent leading order contribution $\langle\sigma^2(x)\rangle\sim m_\sigma^2/(16\pi^2)$, the Ginzburg criterion is still met due to the hierarchy $\Delta_0^2<\kappa^2$.} \cite{Bunch:1978yq,Starobinsky:1994bd},
\begin{align}
	\langle\sigma^2(x)\rangle\sim\left(\frac{H}{m_\sigma}\right)^2\left(\frac{H}{2\pi}\right)^2~.
\end{align}
Thus the left-hand side of (\ref{GinzburgCriterion}) yields
\begin{align}
	\frac{\langle\sigma^2(x)\rangle}{2\mathcal{Z}_\Delta^{00}}\sim\frac{1}{4}\left(\frac{H}{\kappa}\right)^2\left(\frac{H}{\Delta_0}\right)^2\times\Delta_0^2\ll \Delta_0^2~,
\end{align}
where we have applied (\ref{Z00formula}) and the hierarchies $H\ll \kappa,~H\lesssim\Delta_0$. Therefore we conclude that the Ginzburg criterion is indeed met, and that the mean field approximation is valid whenever the gap is allowed to form.


\subsection{Cosmological collider signals}
Now let us turn on the inflaton fluctuations and examine the observational consequences of the collective modes. Our goal is to estimate the resulting curvature bispectrum
\begin{align}
	\langle\zeta_{k_1}\zeta_{k_2}\zeta_{k_3}\rangle'=(2\pi)^4 \mathcal{P}^2_\zeta \frac{1}{(k_1k_2k_3)^2} f_{NL} S(k_1,k_2,k_3)~,
\end{align}
where $\mathcal{P}_\zeta\simeq2\times10^{-9}$ is the curvature power spectrum, $f_{NL}$ is the signal amplitude and $S(k_1,k_2,k_3)$ is the normalized dimensionless shape function. In the cosmological collider paradigm \cite{Chen:2009zp,Baumann:2011nk,Noumi:2012vr,Arkani-Hamed:2015bza}, the squeezed limit of the shape function shows an oscillatory behavior
\begin{align}
	S(k_1,k_1,k_3)\xrightarrow{k_1\gg k_3} \propto \cos\left(\omega\ln\frac{k_1}{k_3}+\cdots\right)~,
\end{align}
where $\omega$ directly gives the mass of the particle exchanged by inflatons, hence cosmological ``collider physics''. 

\begin{figure}[htp] 
	\centering
	\includegraphics[width=10cm]{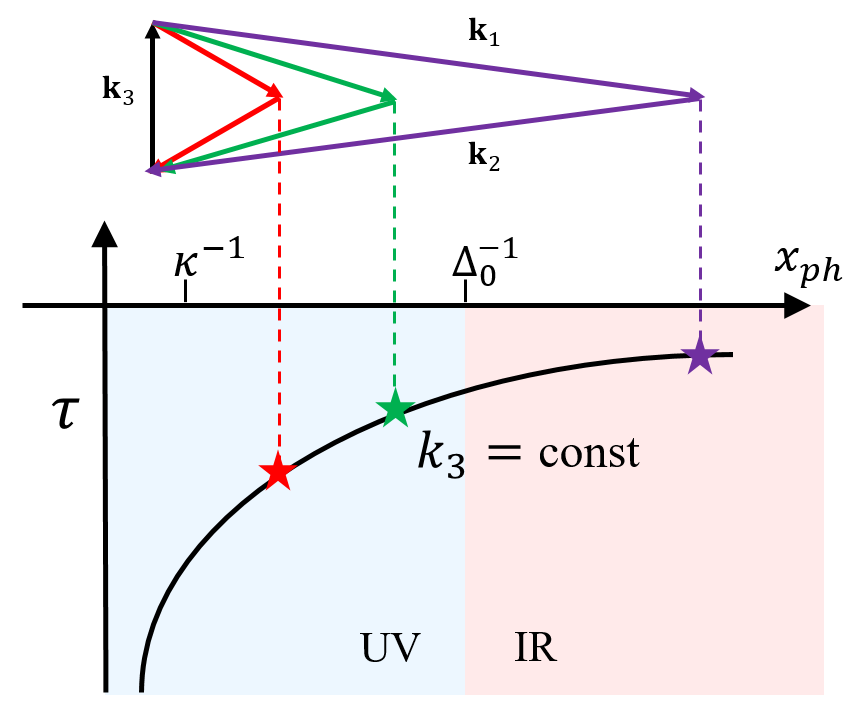}
	\caption{\label{CCschematic} The cosmological collider tomography, i.e. different momentum ratios probe physics at different energy scales. Here the vertical axis is the conformal time $\tau$, and the horizontal axis is the physical distance $x_{\text{ph}}\equiv a(\tau)x=k_{\text{ph}}^{-1}=E^{-1}$. The black curve traces the inflaton mode with a specific comoving momentum $k_3$. The stars correspond to different resonant decay events producing the $\varphi(\mathbf{k}_1),\varphi(\mathbf{k}_2)$ modes. As the universe expands, the energy scale of $\varphi(\mathbf{k}_3)$ decreases, crossing from the UV regime to the IR regime, where it effectively drives different massive degrees of freedom. Now changing the comoving momentum of the other two inflaton modes is equivalent to changing the resonant decay time of the massive degrees of freedom, which is also equivalent to changing the energy scale of the massive degrees of freedom. In this way, the more squeezed the momentum ratio is ($k_1\to \infty$), the later resonance happens, and the lower energy scale is probed by the resonance.}
\end{figure}

As mentioned in Sect.~\ref{flatSpacetimeSect}, our model in the BCS condensed phase can be viewed from two perspectives at different energy scales.
\begin{enumerate}
	\item[$\bullet$] At energy scales above the gap $E>\Delta_0$, it is more natural to take the perspective of a fermionic theory. In this UV picture, the degrees of freedom are carried by interacting fermions with an effective mass generated by the condensate, as described by $S'_f[\psi]$ in (\ref{Sfprime}) with $\langle\Sigma+i\gamma^5\Pi\rangle=\Delta_0 e^{i\gamma^5\Theta_0}$.
	\item[$\bullet$] At energy scales below the gap $E<\Delta_0$, it is more natural to take the perspective of a bosonic theory. In this IR picture, the degrees of freedom are carried by the collective modes, as described by $S_{\text{eff}}[\Delta,\Theta]$.
\end{enumerate}
The inflaton fluctuation in the cosmological collider plays the role of a detector with variable energy scales. This is because in an expanding spacetime, the physical wavelength $E=k_{\text{ph}}=k/a(\tau)$ of an inflaton fluctuation $\varphi(\tau,\mathbf{k})$ is time-dependent. The effective energy scale it probes therefore scans through the theory it couples to. Since physical resonances between the inflaton and different theory particles happen at different energy scales, varying the comoving momentum \textit{ratio}\footnote{It is crucial that only the momentum ratio encodes the physics at different energy scales, since the overall size of the momenta is irrelevant due to the scale invariance.} is therefore tantamount to varying the center-of-mass energy of a particle collider. Hence the cosmological collider can in principle scan through the particle spectrum during inflation, i.e. a tomography in energy scales. In particular, for the inflaton 3-point function, the equilateral limit probes higher-energy physics, while the squeezed limit probes lower-energy physics\footnote{One can also view this cosmological collider tomography from the perspective of the cosmological flow \cite{Werth:2023pfl}. The evolution of the equal-time correlator shows that the formation of cosmological collider signals starts out in the equilateral limit, and extends to the squeezed limit later. This suggests that signals at more squeezed momentum ratios form at lower energy scales due to the redshift.}. This idea is further illustrated by the cartoon diagram in FIG.~\ref{CCschematic}. We shall see this more clearly when we discuss the shape function below.

More explicitly, we need to find the inflaton $\varphi$ coupling to different fields in both the UV picture and the IR picture. The curvature correlator is then translated from inflaton correlators via a gauge transformation
\begin{align}
	\zeta=-\frac{H}{\dot{\phi}_0}\varphi~.
\end{align}	
Naturally, we expect the axial chemical potential to introduce the inflaton coupling via the Stückelberg trick
\begin{align}
	\kappa=\frac{\dot{\phi}_0}{\Lambda_c}\rightarrow\frac{\dot{\phi}_0+\dot{\varphi}(x)}{\Lambda_c}=\kappa+\frac{\dot{\varphi}(x)}{\Lambda_c}~.\label{StückelbergTrick}
\end{align}
Therefore, the inflaton perturbation $\varphi(x)$ couples to fermions in the UV picture via
\begin{align}
	\mathcal{L}_{UV}\equiv\frac{1}{\Lambda_c}\dot{\varphi}\,\bar{\psi}\gamma^0\gamma^5\psi~.
\end{align}
Fermions must form closed loops to contribute to inflaton correlators. In the IR picture, $\varphi(x)$ couples to both the amplitude mode and the phase mode. To find the explicit couplings, we need to expand $S_{\text{eff}}$ in powers of $\varphi$ after doing the Stückelberg trick. We find the leading interactions to be
\begin{align}
	\mathcal{L}_{IR}\equiv \lambda_{\varphi\sigma}\dot\varphi \sigma+\lambda_{\varphi\varphi\sigma}\dot{\varphi}^2\sigma+\left(\lambda_{\varphi\theta\theta}\dot{\varphi}+\lambda_{\varphi\varphi\theta\theta}\dot{\varphi}^2\right)\left(\dot{\theta}^2-\frac{c_\theta^2(\partial_i\theta)^2}{a^2}\right)+\cdots~,
\end{align}
with
\begin{align}
	\lambda_{\varphi\sigma}&=-\frac{\Delta _0 \kappa }{\pi ^2 \Lambda _c\sqrt{2\mathcal{Z}_\Delta^{00}} }\left(\ln \frac{\Delta _0^2}{4 \Lambda ^2}+3\right)\sim\mathcal{O}(1)H,\qquad
	\lambda_{\varphi\varphi\sigma}=\frac{1}{2\kappa\Lambda_c}\lambda_{\varphi\sigma}\sim\mathcal{O}(10^{-4})	\lambda_{\varphi\sigma}/H^2~,\nonumber\\
	\lambda_{\varphi\theta\theta}&=\frac{\kappa }{4 \pi ^2 \Lambda _c\mathcal{Z}_{\Theta }^{00}}\sim\mathcal{O}(10^{-4})H^{-2},\qquad\qquad\qquad\quad\lambda_{\varphi\varphi\theta\theta}=\frac{1}{2\kappa\Lambda_c}\lambda_{\varphi\theta\theta}\sim\mathcal{O}(10^{-4})\lambda_{\varphi\theta\theta}/H^2~,
\end{align}
where we have assumed $M=M_5=40H,\Lambda=\Lambda_\sigma=\Lambda_\theta=100H,\kappa=45H$ for a typical numerical estimate. Notice that by construction (\ref{StückelbergTrick}), the inflaton $\varphi$ is always derivatively coupled. This prevents a tadpole correction for the inflaton background. The phase mode $\theta$ is protected by a $Z_2\subset U(1)_A$ symmetry whether $U(1)_A$ is explicit or not, and always couples to other fields in pairs. This fact brings two implications. First, there can be no mixing between the phase mode with the inflaton. Second, the phase mode contribution to inflaton correlators must starts at loop-level at leading order. The amplitude mode $\sigma$, however, is not protected by any symmetry. It indeed mixes with the inflaton and contributes to the inflaton correlators at tree-level.

\begin{figure}[htp] 
	\centering
	\includegraphics[width=14cm]{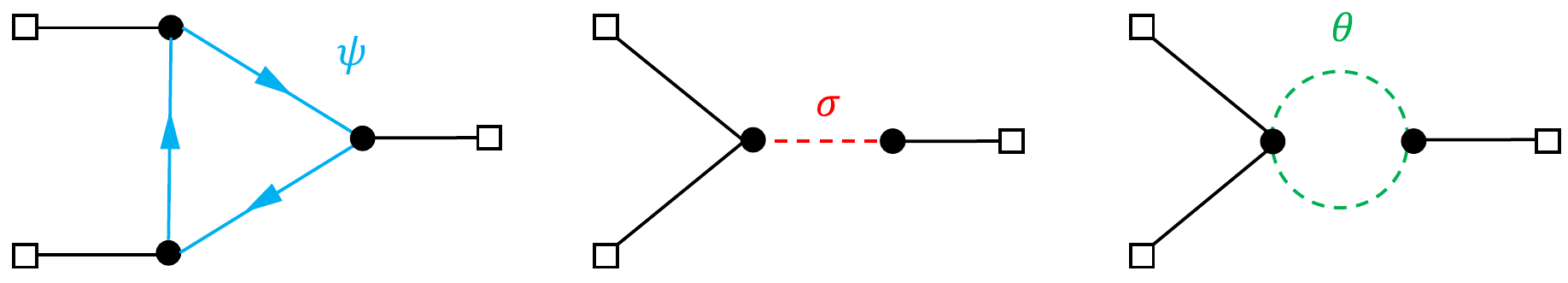}
	\caption{\label{diagram} The leading diagrams that contribute to the bispectrum.}
\end{figure}

The leading order bispectrum comes from a fermion loop in the UV picture, a amplitude mode exchange and phase mode loop in the IR picture (see FIG.~\ref{diagram}).
To estimate the amplitude of signals, we follow the same strategy as in \cite{Wang:2019gbi},
\begin{align}
	f_{\rm{NL}}\sim\frac{1}{2\pi P^{1/2}_\zeta}\times\rm{loop~ factor}\times\rm{vertices}\times\rm{propagators}~,
\end{align}
Propagators are usually suppressed by $1/(\text{mass})$. Additionally, the cosmological collider signals always contain the Boltzmann suppression factor $e^{-\pi (\text{mass})/H}$, attributed to either the resonance saddle point or from the particle vacuum production rates \cite{Tong:2021wai}. The fermion loop contributes in the UV a signal strength \cite{Wang:2019gbi}
\begin{align}
	f^{\psi}_{\rm{NL}}\sim\frac{1}{2\pi P^{1/2}_{\zeta}}\cdot\frac{1}{(4\pi)^2}\cdot\left(\frac{\Delta_0}{\Lambda_c}\right)^3\cdot\frac{4\pi\kappa^2}{H^2}e^{-2\pi(\tilde{\mu}-\tilde{\kappa})}\sim\mathcal{P}_\zeta\Delta_0^3\kappa^5 H^{-8} e^{-\pi\Delta_0^2/\kappa H}\sim\mathcal{O}(1)~,
\end{align}
where the Boltzmann factor is from the particle production rate (\ref{dSfermionOccupationNumber}). In the IR, the amplitude mode gives a negligible tree-level exchange signal
\begin{align}
	f^{\sigma}_{\rm{NL}}\sim\frac{1}{2\pi P^{1/2}_{\zeta}}\cdot\lambda_{\varphi\sigma}\cdot\lambda_{\varphi\varphi\sigma}\cdot\left(\frac{m_\sigma}{H}\right)^{3/2}e^{-\pi m_\sigma/H}\sim \mathcal{O}(10^{-30}).
\end{align}
Here the mass power enhancement is from the three-point vertex, and can be checked through previous works \cite{Chen:2015lza,Tong:2021wai}. The phase mode loop also yields a negligible signal,
\begin{align}
	f^{\theta}_{\rm{NL}}\sim\frac{1}{2\pi P^{1/2}_\zeta}\cdot\frac{1}{(4\pi)^2}\cdot\lambda_{\varphi\theta\theta}\cdot\lambda_{\varphi\varphi\theta\theta}\cdot\left(1+\frac{m_\theta}{H}\right)^{3} e^{-2\pi m_\theta/H}\sim\mathcal{O}(10^{-10})~.
\end{align}
In all the above estimates, we have assumed $M=M_5=40H,\Lambda=\Lambda_\sigma=\Lambda_\theta=100H,\kappa=45H$. Therefore, as we expected, the fermions yield observably large cosmological collider signals, while the collective modes are hardly excited and give essentially no cosmological collider signal. Since different momentum ratios in the squeezed limit probe physics at different energy scales, we expect the oscillatory signal of the UV fermions to \textit{damp out} beyond a scale $k_1/k_3\gtrsim(\cdots)\times\tilde{\Delta}_0^{-1}$.

\begin{figure}[htp] 
	\centering
	\includegraphics[width=13cm]{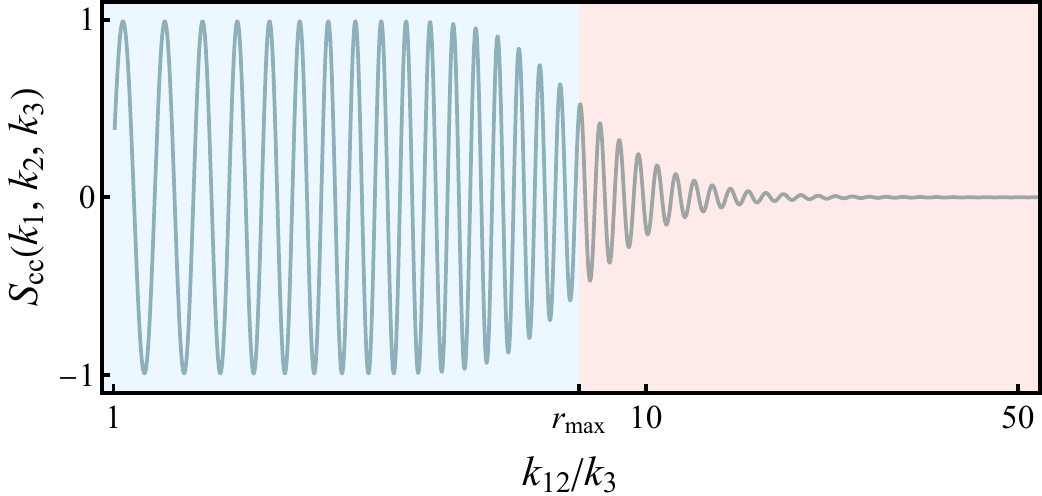}
	\caption{The schematic cosmological collider signal in our BCS model. Here we have chosen the parameters $M=M_5=40H,\Lambda=100H, \kappa=45H$. We have applied an artificial window function centered around $r_{\text{max}}$ to smoothly turn off the fermion signals in the squeezed limit. We caution the readers that this figure is only a rough indication of the features in the true signal, by applying a window function by hand to connect the UV and the IR. We leave a detailed calculation to future works.}\label{CCphaseSchematic}
\end{figure}

To better understand this behavior and find out the waveform of the signal, we will adopt the saddle point method to be developed in details in a forthcoming paper \cite{SPM}. Our only assumption is that the condensation scale $\Delta_0$, serves as a natural boundary between two distinct ``phases''(more precisely, two distinct pictures of the same phase), and that unbounded fermions can only observed in the UV phase where $E\sim \frac{k}{a}>\Delta_0$. First, let us start be noticing that cosmological collider signals measure the total phase cumulated between different events~\cite{Tong:2021wai}. For fermions with a comoving momentum close to that of $\varphi({\mathbf{k}_3})$ in the UV, the first relevant event is its particle production, at a time $\tau_*$ given by \cite{Sou:2021juh}
\begin{align}
	\text{Production:} \qquad	-k_3\tau_*=0.66\tilde{\mu}+0.34\tilde{\kappa}\approx \tilde{\kappa}~,
\end{align}
The UV fermions then propagate and interact, before condensing at a time $\tau_c$ with 
\begin{align}
	\text{Condensation:} \qquad	-k_3\tau_c=\tilde{\Delta}_0~.
\end{align}
The last possible event is the resonant decay of fermions into inflaton pairs with momenta $\mathbf{k}_1,\mathbf{k}_2$. To determine more precisely the resonance time $\tau_\bullet$, we note that the dynamical phase of the fermion mode function takes the form $e^{-i\int w(\tau)d\tau}$ with dispersion relation $w^2=(k_3-\tilde{\kappa} a)^2+\tilde{\Delta}_0^2a^2$, while the inflaton oscillates as $e^{i k_{12} \tau}$. Then the resonance event happens when the total dynamical phase strikes a saddle point,
\begin{align}
	\text{Resonance:} \qquad\frac{d}{d\tau}\arg \left[ e^{i k_{12} \tau}\times e^{-2i\int w(\tau)d\tau}\right]\Bigg|_{\tau_\bullet}=k_{12}-2\sqrt{(k_3+\tilde{\kappa}/\tau_\bullet)^2+\tilde{\Delta}_0^2/\tau_\bullet^2}=0~.
\end{align}
The factor of two in the fermion dynamical phase arises from the loop where two internal fermion propagators are soft. We have also approximated the hard fermion in the loop as an EFT contact term. The cosmological collider signal records the cumulated phase from the production and the resonance, while the condensation time sets a cutoff for the latest resonance. Therefore, the necessary condition for the existence of the cosmological collider signal is the time hierarchy
\begin{align}
	\tau_c>\tau_\bullet>\tau_*~.
\end{align}
This gives a momentum ratio cutoff for the cosmological collider signal,
\begin{align}
	\frac{k_{12}}{k_3}<r_\text{max}\equiv\frac{2\kappa}{\Delta_0}\sqrt{1-\frac{2\Delta_0}{\kappa}+\frac{2\Delta_0^2}{\kappa^2}}~.
\end{align}
For a momentum ratio beyond this cutoff scale, the resonance event cannot happen because the fermions would have paired and condensed due to the attractive self-interaction. The exact expression for the correlator is hard to calculate even at the tree-level \cite{Arkani-Hamed:2018kmz,Qin:2022fbv,Qin:2023ejc} or for a bubble loop diagram \cite{Xianyu:2022jwk}, not to mention the triangle loop diagram here. Thus here we simply estimate the cosmological collider signal by the cumulated phase 
\begin{align}
	S_{cc}(k_1,k_2,k_3)\sim& \cos \left[k_{12}\tau_\bullet-2\int^{\tau_\bullet}d\tau\sqrt{\left(k_3+\frac{\tilde{\kappa}}{\tau}\right)^2+\frac{\tilde{\Delta}_0^2}{\tau^2}}\right]\nonumber\\
	=&\cos\left[4\tilde{\mu}\,\mathrm{arctanh}\,\frac{2\tilde{\mu}}{\left(k_{12}+2k_3\right)\tau}-\tilde{\kappa}\ln\frac{4 k_3}{{\left[4\tilde{\kappa}+\left(k_{12}+2k_3\right)\tau_\bullet\right]^2}}\right]~,\quad \frac{k_{12}}{k_3}<r_\text{max}~.
\end{align} 
We plot this approximate waveform in FIG.~\ref{CCphaseSchematic} with a schematic window function to indicate the damping effect beyond $r_\text{max}$. We stress that this is can only be understood as a very rough description of what the true signal might be, due to the various approximations made above. A more refined computation is left for future works.


\subsection{Gravitational waves from phase transition}
The time dependence of the chemical potential $\kappa=\dot{\phi}_0(t)/\Lambda_c$ directly enters the shape of the effective potential. Therefore, if the fermion theory starts out in the normal phase with $\langle\bar{\psi}\psi\rangle=0$, the acceleration of the rolling inflaton may trigger a first-order phase transition that is associated with BCS condensation $\langle\bar{\psi}\psi\rangle=2M^2\Delta_0$. The bubble collisions during the phase transition then become the source of stochastic gravitational waves. To estimate the gravitational wave energy density today, we follow the treatment of \cite{An:2020fff,An:2022cce,An:2022toi} and examine the changing rate of the bubble Euclidean action. In the Euclidean IR effective theory, the leading order action reads
\begin{align}
	S_4=\int dtd^3x\left[\mathcal{Z}^{00}_\Delta\dot\Delta^2-\mathcal{Z}^{33}_\Delta (\partial_i \Delta)^2+V_{\mathrm{eff}}(\Delta)\right]~.
\end{align}
Before solving for the bounce solution, we canonicalize the gap field by introducing
\begin{align}
	\rho\equiv \int\sqrt{2 \mathcal{Z}^{00}_\Delta} d\Delta\approx \frac{\kappa}{\sqrt{6}\pi}\ln\frac{\Delta}{\Delta_{r}}~,
\end{align}
and rewrite the action as
\begin{align}
	S_4\approx\int dtd^3x\left[\frac{1}{2}\dot\rho^2+\frac{1}{2}c_s^2 (\partial_i \rho)^2+V_{\mathrm{eff}}(\Delta(\rho))\right]~.
\end{align}
with $c_\sigma^2\equiv -\mathcal{Z}^{00}_\Delta/\mathcal{Z}^{33}_\theta\approx 1/3$ and $\Delta_{r}$ being an arbitrary reference scale. To solve for an $O(4)$-invariant bounce, we rescale the spatial coordinates by defining $\bar{x}\equiv c_s x$ and absorb the non-unit sound speed,
\begin{align}
	S_4\approx\frac{1}{c_s^3}\bar{S}_4\quad,\quad \bar{S}_4\equiv \int dtd^3\bar{x}\left[\frac{1}{2}\dot\rho^2+\frac{1}{2} (\bar{\partial}_i \rho)^2+V_{\mathrm{eff}}(\Delta(\rho))\right]~.
\end{align}
Now $\bar{S}_4$ gives the $O(4)$-invariant bounce action in the thin-wall limit as
\begin{align}
	\bar{S}_4=\frac{27\pi^2 \bar{S}_1^4}{2{\delta V_{\mathrm{eff}}}^3}~.
\end{align}
Here
\begin{align}
	\bar{S}_1=\int_{-\infty}^{\rho_0}\sqrt{2V_{\mathrm{eff}}(\Delta(\rho))}d\rho~,
\end{align}
is the one-dimensional instanton action and $\delta V_{\mathrm{eff}}$ is the vacuum energy difference.	During the phase transition, the typical change of vacuum energy can be estimated as
\begin{align}
	\delta V_{\mathrm{eff}}\approx\frac{\kappa^2\Delta_0^2}{2\pi^2}~.
\end{align}
The $\rho$ field traverse is typically $\delta \rho=\mathcal{O}(1)\kappa/\sqrt{6}\pi$. Therefore, we approximate
\begin{align}
	\bar{S}_1\approx \sqrt{2\delta V_{\mathrm{eff}}}\times\frac{\kappa}{\sqrt{6}\pi}=\frac{\kappa^2\Delta_0}{\sqrt{6}\pi^2}~.
\end{align}
This leads to an estimated bounce action
\begin{align}
	S_4\approx \frac{\kappa^2}{\sqrt{3}\Delta_0^2(\kappa)}~,
\end{align}
and its time derivative
\begin{align}
	\frac{\beta}{H}\equiv-\frac{dS_4}{Hdt}=-\frac{\partial S_4}{\partial\ln\kappa}\frac{\ddot{\phi}_0}{H\dot\phi_0} \approx-\frac{\eta}{2}\frac{\partial S_4}{\partial\ln\kappa}\sim\eta\left(\bar{M}^2\pi^2-\kappa^2\right)\frac{e^{\frac{\bar{M}^2\pi^2}{\kappa^2}}}{4\sqrt{3}\Lambda^2},
\end{align}
where $\eta$ is the second slow-roll parameter. 
The spectrum of gravitational waves today can be written as \cite{An:2020fff,An:2022cce,An:2022toi}
\begin{align}
	\Omega_{\rm{GW}}(f)=\Omega_R\times\mathcal{S}(2\pi f)\times\frac{\delta V_{\mathrm{eff}}}{\rho_{\rm{\inf}}}\times\frac{\mathrm{d}\rho^{\rm{flat}}_{\rm{GW}}}{\delta V_{\mathrm{eff}}\,\mathrm{d}\ln f}~.
\end{align}
Here $\Omega_R$ is the present-day radiation abundance, while $\rho_{\rm{\inf}}$ is the inflaton potential energy density. The definition of $\mathcal{S}$  is provided in \cite{An:2020fff} and the flat space spectrum is given by \cite{Huber:2008hg}
\begin{align}
	\frac{\mathrm{d}\rho^{\rm{flat}}_{\rm{GW}}}{\delta V_{\mathrm{eff}}\,\mathrm{d}\ln f}=\left(\frac{\delta V_{\mathrm{eff}}}{\rho_{\rm{inf}}}\right)\times\left(\frac{H}{\beta}\right)^2\times\mathrm{W}\times\frac{3.8k_{*}\left(2\pi f\right)^{2.8}}{k_{*}^{3.8}+2.8\left(2\pi f\right)^{3.8}}~,
\end{align}
with $\mathrm{W}\sim 0.077$ and $k_{*}\sim1.44\beta$~. The ratio between the latent heat and the inflaton potential energy density is estimated in (\ref{LatentHeatFractionEstimate}) as
\begin{align}
	\delta\rho_r\equiv\frac{\delta V_{\mathrm{eff}}}{\rho_{\rm{inf}}}\approx\frac{\kappa^2\Delta_0^2}{6\pi^2M_{\rm{pl}}^2H^2}\lesssim\frac{\varepsilon}{12\pi^2}~.
\end{align}
Despite the constraints imposed by experimental data on the magnitude of the slow-roll parameter at CMB scales, it is worth noting that the generation of gravitational waves during inflation may occur in later stages of the inflation, where the slow-roll parameter $\varepsilon$ may take a relatively larger value (e.g. $\varepsilon\sim 0.1$), then the density ratio $\delta\rho_r$ can be as large as $\mathcal{O}(10^{-3})$~.
\begin{figure}[htp]
	\centering
	\includegraphics[width=15cm]{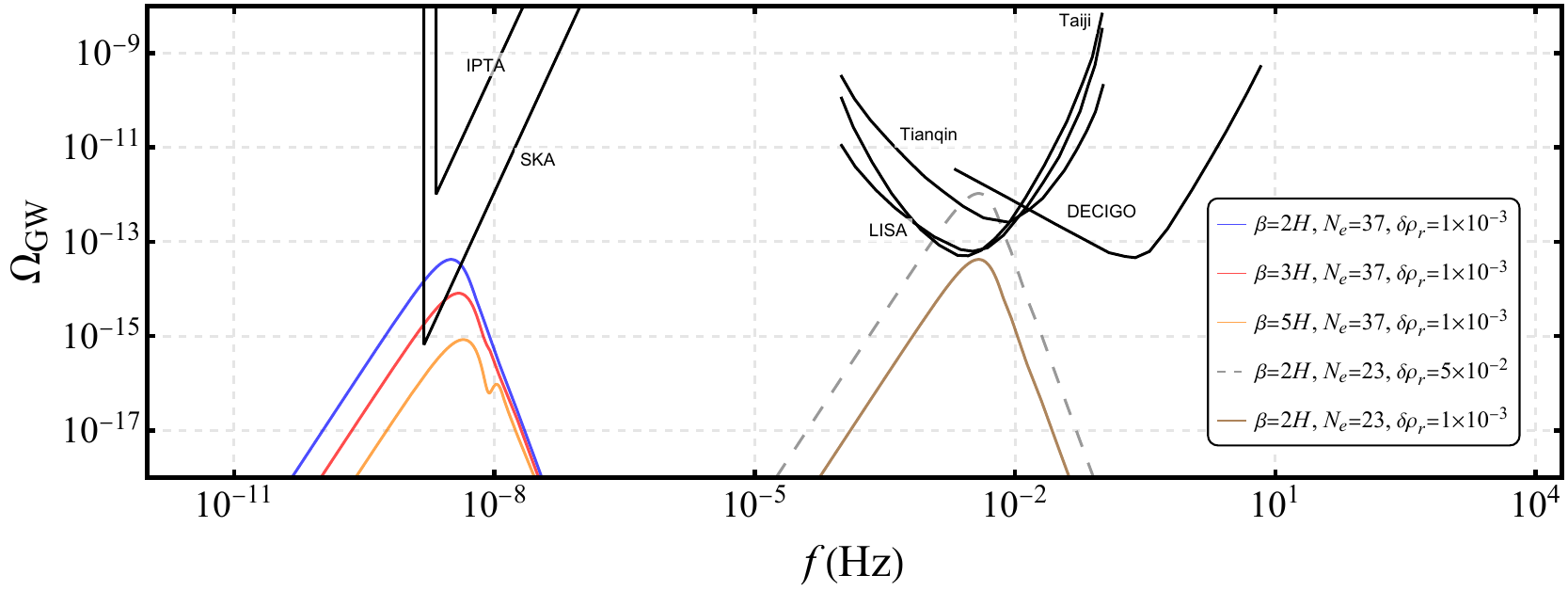}
	\caption{\label{GW} The gravitational waves energy density $\Omega_{\rm{GW}}$ from BCS phase transition as a function of frequency $f$. The legend insertion of the figure explicitly displays the various parameter selections, where $N_e$ is the e-folding number before the end of inflation. The experiment curves are from IPTA~\cite{Hobbs:2009yy}, SKA~\cite{Janssen:2014dka}, as well as space-based detectors such as LISA~\cite{LISA:2017pwj}, DECIGO~\cite{Kawamura:2006up}, TianQin\cite{TianQin:2015yph}, Taiji~\cite{Ruan:2018tsw}. The gray line is dashed to indicate a risky parameter choice $\varepsilon=0.5$. Unfortunately, for $\varepsilon\lesssim 0.1$ in general, we do not expect the signal strength (brown solid line) to reach the sensitivity of LISA or DECIGO.}
\end{figure}

We plot the resulting gravitational waves signals in the Fig.\ref{GW}. The e-folding number is chosen as $N_e=37$ for the frequency region relevant to Pulsar Timing Array (PTA) experiments, whereas for space-based detectors such as LISA, it is set to $N_e=23$. We see that the gravitational waves associated with the BCS phase transition is feasible for the SKA experiment, while the detection probability for space-based detectors is less optimistic, unless the parameters are tuned to risky values such as $\varepsilon \sim 0.5$.

\section{Conclusion}\label{ConclusionSect} 
The BCS theory of superconductivity has long been a thriving paradigm that inspired many excellent works and exciting discoveries across the vast territories of modern physics. It is particularly interesting to consider its generalization to cosmology due to the similarity between the expanding universe and condensed matter systems. In this work, we considered embedding a class of BCS-like models with chemical potential into inflation, and studied in detail the properties of the fermion condensate and its phenomenology. Starting from the single-coupling case, we computed analytically the effective potential for BCS condensation in dS non-perturbatively in the spacetime curvature, subject to the decoupling limit $M_p\to \infty$ and neglecting backreaction from the inflaton and graviton sector. We found that the spacetime curvature effect strongly impacts the fermion condensation. Namely, the spacetime expansion may rip Cooper pairs apart and forbid the formation of an exponentially small gap as would otherwise be present in the flat spacetime approximation. During the slow-roll phase of inflation, BCS condensate can form with a gap at least a few times the Hubble scale, with the assistance of chemical potential. Assuming some reasonable constraints, we were able to prove that the corresponding BCS phase transition is always first-order, and obtained the phase diagram of our system. Most features remain the same after we generalize to the two-coupling model, where an additional phase mode appears in the IR EFT. The analysis of the collective modes shows that they share a universal sound speed of relativistic fluids, and masses proportional to the gap. We checked the Ginzburg criterion and showed the validity of the mean-field approximation. Our BCS model in the sky also yields interesting (although challenging at the same time) observational signatures via the cosmological collider tomography or stochastic gravitational waves induced by the first-order BCS phase transition. There are certainly many aspects for further investigations in the future:
\begin{enumerate}
	\item[$\bullet$] Throughout this work, we have limited ourselves to the adiabatic case with a slowly-varying yet non-zero Hubble parameter. Such is the case for inflation. However, in later stages of the universe, the Hubble parameter is itself time-dependent. The adiabatic result with an instantaneous Hubble parameter can only serve as a crude leading order approximation. It would be interesting to give a better analysis of BCS condensations in a general FRW spacetime with significant deviations from dS.
	\item[$\bullet$] As mentioned in Sect.~\ref{U1A}, we have excluded the simplest vector condensate model due to the spatial isotropy. However, in more complicated cases where the internal symmetries mix with spacetime symmetries, it is possible to form a vector condensate while maintaining the (effective) spatial isotropy. We leave this interesting scenario for future studies.
	\item[$\bullet$] For simplicity, we only focused on the essential part of the BCS model in this work, namely its pairing of fermions into Cooper pairs and their condensation. The gauge fields have not yet been introduced, and there is no Higgs phase of superconductivity. Hence our model in this work resembles more that of a fermionic superfluid. It is thus interesting to introduce charged fermion condensate and gauge fields along with their spontaneous symmetry breaking, thereby achieving a cosmological superconducting phase. This is left for future work.
	\item[$\bullet$] Our proof for the BCS phase transition being first-order is made under a naturalness assumption. This is reasonable from a practical point of view, yet it may also be possible to relax such constraints and check the transition type in more general cases. In other words, if we relax certain constraints, are there any critical points where the phase boundary terminates, and a second-order phase transition takes over?
	\item[$\bullet$] Despite the arguments built upon strong physical intuitions, we did not give any precise computation of the cosmological collider signal and its attenuation behavior when crossing into the IR regime. Thus an important question for us to work on in the future is how to acquire a more quantitative grasp on this tomography of cosmological colliders.
\end{enumerate}


\acknowledgments
We would like to thank Yanbai Gao, Sida Lu and Chon Man Sou for helpful discussions. This work was supported in part by the National Key R\&D Program of China (2021YFC2203100), the NSFC Excellent Young Scientist Scheme (Hong Kong and Macau) Grant No. 12022516, and by the RGC of Hong Kong SAR, China Grant No. 16306422. XT is supported in part by STFC consolidated grants ST/T000694/1 and ST/X000664/1. YZ is also supported by the IBS under the project code, IBS-R018-D3.


\appendix

\section{The dS loop integral}\label{dSloopApp}
Let us first consider the asymptotic behaviors of above integrals in the limit $z\rightarrow\infty$. The product of Whittaker functions admits an asymptotic expansion
\begin{align}
	W_{\frac{1}{2}-i\tilde{\kappa},i\tilde{\mu}}(-2iz)\,W_{-\frac{1}{2}+i\tilde{\kappa},-i\tilde{\mu}}(2iz)=-i e^{-\pi  \tilde{\kappa }} \left(1+\frac{\tilde{\kappa }+\frac{i}{2}}{z}+\frac{3 \tilde{\kappa }^2+3 i \tilde{\kappa }-\tilde{\mu }^2-1}{2 z^2}+\mathcal{O}(z^{-3})\right)~.
\end{align}
The leading term leads to a quadratic divergence, as expected from the flat spacetime computation. The $\mathcal{O}(z^{-1})$ term leads to a superficial linear divergence which is canceled out after summing over helicities. The $\mathcal{O}(z^{-2})$ term gives rise to a logarithmic divergence similar to that in flat spacetime, but with an extra correction due to spacetime curvature. The rest of the $\mathcal{O}(z^{-3})$ terms are convergent for the loop integral but diverge as an infinite series. The above inspection of degree of divergence is helpful because it allows us to extract all the cutoff-dependent contributions in the loop integral. For instance, cutting off the integral $\mathcal{I}_-$ at $\Lambda/H\gg1$ with $\Lambda\lesssim \pi M$, we have
\begin{align}
	\mathcal{I}_-(\Lambda)=i\int_{0}^{\Lambda/H}dz\,z W_{\frac{1}{2}-i\tilde{\kappa},i\tilde{\mu}}(-2iz)\,W_{-\frac{1}{2}+i\tilde{\kappa},-i\tilde{\mu}}(2iz)~,
\end{align}
Then we take a derivative of $\Lambda$ and expand in inverse powers of the cutoff:
\begin{align}
	e^{\pi\tilde{\kappa}}\frac{\partial \mathcal{I}_-(\Lambda)}{\partial\Lambda}&=i \,e^{\pi\tilde{\kappa}}\frac{\Lambda}{H^2}W_{\frac{1}{2}-i\tilde{\kappa},i\tilde{\mu}}(-2i\Lambda/H)\,W_{-\frac{1}{2}+i\tilde{\kappa},-i\tilde{\mu}}(2i\Lambda/H)\nonumber\\
	&=\frac{\Lambda}{H^2}+\left(\frac{\kappa}{H^2}+\frac{i}{2H}\right)+\left(\frac{\kappa^2-\Delta^2/2}{H^2}+\frac{3i\kappa}{2H^2}-\frac{1}{2}\right)\frac{1}{\Lambda}+\mathcal{O}\left(\frac{1}{\Lambda^2}\right)~.
\end{align}
Integrating it back to $\mathcal{I}_-(\Lambda)$, we recover all the cutoff-dependent terms, plus an unknown finite remainder that is independent of the cutoff,
\begin{align}
	e^{\pi\tilde{\kappa}}\mathcal{I}_-(\Lambda)=\frac{\Lambda^2}{2H^2}+\left(\frac{\kappa}{H^2}+\frac{i}{2H}\right)\Lambda+\left(\frac{\kappa^2-\Delta^2/2}{H^2}+\frac{3i\kappa}{2H^2}-\frac{1}{2}\right)\ln\frac{\Lambda}{H}+\mathcal{O}\left(\frac{1}{\Lambda}\right)+f^\Lambda_-(\Delta,\kappa,H)~,\label{LambdaRegIminus}
\end{align}
where the superscript on $f^\Lambda_-(\Delta,\kappa,H)$ denotes the regularization scheme ($\Lambda$-reg), not the functional dependence. Such a finite remainder is the truly non-trivial part of the loop integral in dS, and characterizes the curved spacetime effect on the behavior of the system. We need to find its explicit form. Summing over helicities, we have
\begin{align}
	\frac{\partial V_{\rm{1-loop}}}{\partial \Delta} =-\frac{\Delta \Lambda^2}{2\pi^2}-\left(\frac{\kappa^2-H^2/2}{\pi^2}\Delta-\frac{\Delta^3}{2\pi^2}\right)\ln \frac{\Lambda}{H}+\mathcal{O}(\Lambda^{-2})-\frac{\Delta H^2}{4\pi^2}\left[f^{\Lambda}_++f^{\Lambda}_-+\text{c.c.}\right]~.
\end{align}

Like the flat-spacetime QFT, there are multiple ways of regularizing the integral, and they must match each other up to some $\Delta$-independent constants.	Unfortunately, it turns out to be challenging to compute $f^\Lambda_{\pm}$ directly in the $\Lambda$-reg scheme. Thus, we seek to use another regularization scheme ($\epsilon$-reg) to compute the finite term $f^{\epsilon}_\pm(\Delta,\kappa,H)$ and match them by allowing a difference degenerate with the two divergences:
\begin{align}
	f^{\Lambda}_{\pm}(\Delta,\kappa,H)-f^{\epsilon}_{\pm}(\Delta,\kappa,H)=A(\kappa,H)+B(\kappa,H)\Delta^2/H^2~.\label{matchingCondition}
\end{align}
so that after renormalization, they give the same result. The $\epsilon$-reg scheme is possible thanks to the Whittaker product integral formula of Arthur Erdelyi ~\cite{Gradshteyn:1702455},
\begin{align}
	\int_{0}^{\infty}x^{\rho-1}e^{-\frac 12 (\alpha+\beta)x}W_{\kappa,\mu}(\alpha x)W_{\lambda,\nu}(\beta x)dx=\beta^{-\rho}\,G^{22}_{33}\,\Bigg(\frac{\beta}{\alpha}\Bigg|\,\begin{array}{c} \frac{1}{2}+\mu, \frac{1}{2}-\nu,1-\lambda+\rho\\[2.5pt] \frac{1}{2}+\mu+\rho, \frac{1}{2}-\nu+\rho,\kappa \end{array}\Bigg)~,
\end{align}
where $G^{22}_{33}$ is the Meijer $G$-function \cite{NIST:DLMF}. To proceed, we deform $\mathcal{I}_-$ by
\begin{align}
	\mathcal{I}_-(\epsilon)&=i \int_{0}^{\infty}dz\,z\, e^{i(1-r)z}W_{\frac 12-i\tilde{\kappa},i\tilde{\mu}}(-2iz)\,W_{-\frac 12+i\tilde{\kappa},-i\tilde{\mu}}(2irz)\nonumber\\
	&=-\frac{i}{4r^2}\,G^{22}_{33}\,\Bigg(-r\Bigg|\,\begin{array}{c} \frac{1}{2}-i\tilde{\mu}, \frac{1}{2}+i\tilde{\mu},\frac{7}{2}-i\kappa\\[2.5pt] \frac{5}{2}-i\tilde{\mu}, \frac{5}{2}+i\tilde{\mu}+\rho,\frac 12-i\kappa \end{array}\Bigg)~,
\end{align}
with $r=1-i\epsilon$. The Meijer $G$-function is reducible to combinations of the generalized hypergeometric function, which is singular at unit argument where $\epsilon=0$,
\begin{align}
	G_{33}^{22}\left(-r\left|
	\begin{array}{c}
	a_1,a_2,a_3 \\
	b_1,b_2,b_3 \\
	\end{array}
	\right.\right)=&\nonumber \mathcal{A}\, x^{b_1}\,{}_3\mathrm{F}_2\left(\begin{matrix}
	b_1-a_1+1,b_1-a_2+1,b_1-a_3+1\\
	b_1-b_2+1,b_1-b_3+1
	\end{matrix}\middle|r\right)+\big(b_1\leftrightarrow b_2\big)~,
\end{align}
with
\begin{align}
	\mathcal{A}=\frac{\pi  \csc \pi  \left(b_2-b_1\right) \Gamma \left(b_1-a_1+1\right) \Gamma \left(b_1-a_2+1\right)}{\Gamma \left(b_1-b_2+1\right) \Gamma \left(b_1-b_3+1\right)
		\Gamma \left(a_3-b_1\right)}~.
\end{align}
The expansion of generalized hypergeometric function at unit argument is \cite{hyper}
\begin{align}
	&\nonumber\frac{\Gamma(a_1)\Gamma(a_2)\cdots\Gamma(a_{p+1})}{\Gamma(b_1)\cdots\Gamma(b_{p})}
	{}_{p+1}\mathrm{F}_p\left(\begin{matrix}
	a_1,a_2,\cdots,a_{p+1}\\
	b_1,\cdots,b_p
	\end{matrix}\middle|z\right)\\
	&=(1-z)^{-t}\sum^{t-1}_{n=0}h_n(1-z)^n+\sum^{\infty}_{n=0}\left[u_n+v_n\ln (1-z)\right](1-z)^n~,
\end{align}
which is valid if $s=\sum^{p}_{i=1}b_i-\sum^{p+1}_{i=1}a_i$ is equal to an integer $t$ such that $-t\le0$. For a detailed expression of all types of coefficients, we refer readers to \cite{hyper}. Applying the above formula one can write 
\begin{align}
	e^{\pi\tilde{\kappa}}\mathcal{I}_-(\epsilon)=\frac{1}{4\epsilon^2}+\frac{3}{4\epsilon}\left(\frac{\kappa}{H}+\frac{i}{2}\right)-\left(\frac{\kappa^2-\Delta^2/2}{H^2}+\frac{3i\kappa}{H}-\frac{1}{2}\right)\ln \epsilon+\mathcal{O}(\epsilon)+f^{\epsilon}_-(\Delta,\kappa,H)~.\label{epsRegIminus}
\end{align}
Notice that now the finite remainder $f^{\epsilon}_-(\Delta,\kappa,H)$ is a known function. One can also compare (\ref{epsRegIminus}) with (\ref{LambdaRegIminus}) and see that the divergence structure is similar. In particular, the coefficients for the logarithmic divergence are identical. Repeating the procedure for $f^{\epsilon}_+$ and sum over helicities, we obtain
\begin{align}
	\frac{\partial V_{\rm{1-loop}}}{\partial \Delta} =-\frac{\Delta H^2}{4\pi^2\epsilon^2}+\left(\frac{\kappa^2-H^2/2}{\pi^2}\Delta-\frac{\Delta^3}{2\pi^2}\right)\ln \epsilon+\mathcal{O}(\epsilon)-\frac{\Delta H^2}{4\pi^2}\left[f^{\epsilon}_++f^{\epsilon}_-+\text{c.c.}\right]~.
\end{align}
where the last term enjoys an exact analytical expression,
\begin{keyeqn}
	\begin{align}
		\Re\left[f^{\epsilon}_-+f^{\epsilon}_+\right]=&\,\frac{1}{{16 \left(\tilde{\kappa}^2 -\tilde{\mu}^2 \right) }}\bigg\{\tilde{\mu} ^2\left[4 (7+16 \gamma ) \tilde{\kappa} ^2-24 \pi  \tilde{\kappa} -16 \gamma -15\right] -16 \gamma  \tilde{\mu} ^4\nonumber\\
		&~~~~~~~~~~~~~~~~~~~~~~+24 \tilde{\kappa}  \tilde{\mu}  \bigg[\csch(2\pi\tilde{\mu}) \left(\big(2 \tilde{\kappa} ^2-2 \tilde{\mu} ^2-1\big) \sinh (2 \pi  \tilde{\kappa} )+\cosh
		(2 \pi  \tilde{\kappa} )\right)-\coth(2\pi\tilde{\mu}) \bigg]\nonumber\\
		\nonumber&~~~~~~~~~~~~~~~~~~~~~~+\tilde{\kappa} ^2 \left[ 24 \pi  \tilde{\kappa} +16 \gamma +39-4 (7+12 \gamma )\tilde{\kappa} ^2\right]\bigg\}\\
		\nonumber&+\text{csch}(2\pi\tilde{\mu})\bigg\{ \sinh\pi  (\tilde{\kappa}+\tilde{\mu}) \psi (i \tilde{\kappa} +i \tilde{\mu} )\bigg[\left(\tilde{\mu} ^2-3 \tilde{\kappa} ^2+1\right) \cosh \pi(\tilde{\kappa}-\tilde{\mu} )-3i\tilde{\kappa} \sinh \pi  (\tilde{\kappa} -\tilde{\mu}
		)\bigg]\\
		&~~~~~~~~~~~~~~~~~~~~~~+\big(\tilde{\mu}\rightarrow-\tilde{\mu}\big)+\text{c.c.}\bigg\}~.\label{finiteRemainder}
	\end{align}
\end{keyeqn}
Here $\gamma$ is the Euler constant and $\psi(z)=\Gamma'(z)/\Gamma(z)$ is the digamma function. As we have mentioned before, different regularization methods should yield the same physical result after renormalization. Any constant-mismatch before renormalization should be regarded as the mismatch of divergent terms. To match with the $\Lambda$-reg result, we require (\ref{matchingCondition}) and set
\begin{align}
	\frac{\partial V_{\rm{1-loop}}}{\partial \Delta} =&-\frac{\Delta \Lambda^2}{2\pi^2}-\left(\frac{\kappa^2-H^2/2}{\pi^2}\Delta-\frac{\Delta^3}{2\pi^2}\right)\ln \frac{\Lambda}{H}+\mathcal{O}(\Lambda^{-2})-\frac{\Delta H^2}{4\pi^2}\left[f^{\epsilon}_++f^{\epsilon}_-+2A+\frac{2B\Delta^2}{H^2}+\text{c.c.}\right]~.\label{beforeAB}
\end{align}
for some chosen $\Delta$-independent functions $A(\kappa,H)$, $B(\Lambda,H)$. To determine the functions $A,B$, we compare the flat spacetime $\Lambda$-reg result (\ref{flatV}) with (\ref{beforeAB}) expanded up to $\mathcal{O}(H^0)$, and demand the equality for an arbitrary $\Delta$. This gives two equations that determine the unknown functions as
\begin{keyeqn}
	\begin{align}
		A(\kappa,H)=\frac{\kappa ^2}{H^2}\left(\gamma +\ln 2-\frac{5}{8}\right)~,\qquad B(\kappa,H)=\frac{1}{4}(1-2 \gamma -2 \ln 2)~.\label{ABcoefficientFunctions}
	\end{align}
\end{keyeqn}

\begin{figure}[htp] 
   	\centering
   	\includegraphics[width=10cm]{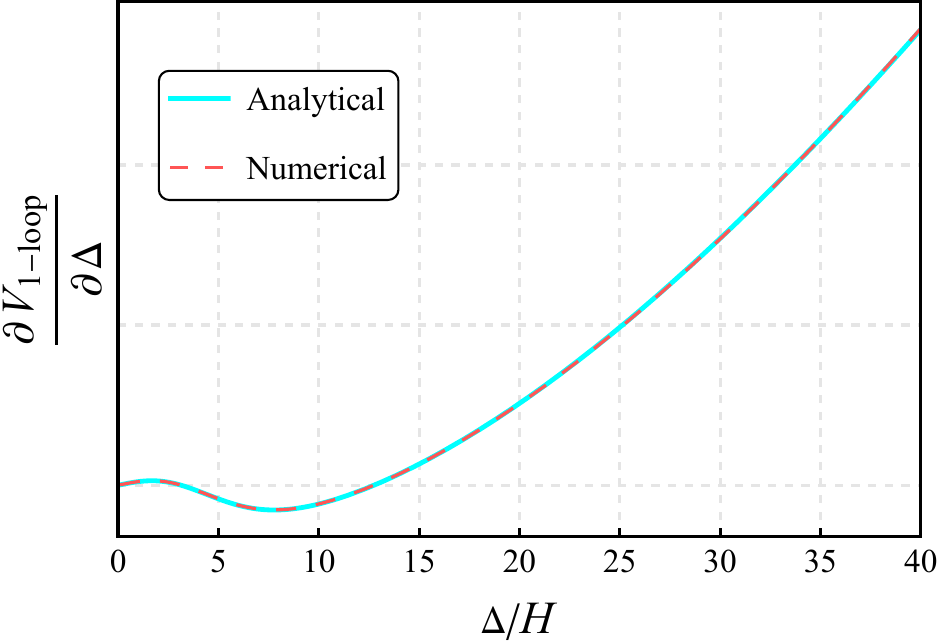}
   	\caption{\label{numVSana} The comparison between the analytical expression (cyan solid) of $\partial V_{\rm{1-loop}}/{\partial \Delta}$ and the numerical calculation (red dashed). The parameters are chosen to be $M=40H,\Lambda=100H,\kappa=50H$ here.}
\end{figure}
As a small caveat, this only recovers the leading order of $A,B$ in $H$. But practically it does not matter, because one always needs to perform the renormalization. In addition, it does not influence the shape of the effective potential by much. One could have started with $\epsilon$-reg directly. We prefer the $\Lambda$-reg only because it is conceptually simple and convenient for the numerical check.
In summary, our exact result for the effective potential is
\begin{align}
	\nonumber V_{\rm{eff}}(\Delta)&=\frac{1}{4}M^2\Delta^2+\int d\Delta 	\frac{\partial}{\partial \Delta} V_{\rm{1-loop}}\nonumber\\
	\nonumber&=\frac{1}{4}\left(M^2-\frac{\Lambda^2}{\pi^2}\right)\Delta^2-\left(\frac{\kappa^2-H^2/2}{2\pi^2}\Delta^2-\frac{\Delta^4}{4\pi^2}\right)\ln\frac{\Lambda}{H} -\int d\Delta \frac{\Delta H^2}{2\pi^2}\Re\left[f^{\epsilon}_++f^{\epsilon}_- +2A+\frac{2B\Delta^2}{H^2}\right]~.
\end{align}
To confirm its validity, we numerically compute the loop integrals, and find the result being consistent with our analytical result, as shown in FIG.~\ref{numVSana}.


\section{Field strength renormalization}\label{ZloopApp}
In order to calculate the 1-loop diagrams in (\ref{Zcoeff}), we first determine the fermion propagator with chemical potential, which can be obtained by inverting its equation-of-motion,
\begin{align}
	S(p)&=\frac{-i}{\slashed{p}-\Delta+\kappa \gamma^0\gamma^5}=i\frac{\left(\slashed{p}+\Delta-\kappa\gamma^5\gamma^0\right)\left[p^2+\Delta^2+\kappa^2-\kappa\{\slashed{p},\gamma^5\gamma^0\}\right]}{\left(p^2+\Delta^2+\kappa^2\right)^2-4\kappa^2\left(p_0^2+p^2\right)}~.
\end{align}
Then the 1-loop field strength renormalization of $\sigma$ can be expressed as following integral
\begin{align}
	\parbox{0.25\textwidth}{\includegraphics[width=0.25\textwidth]{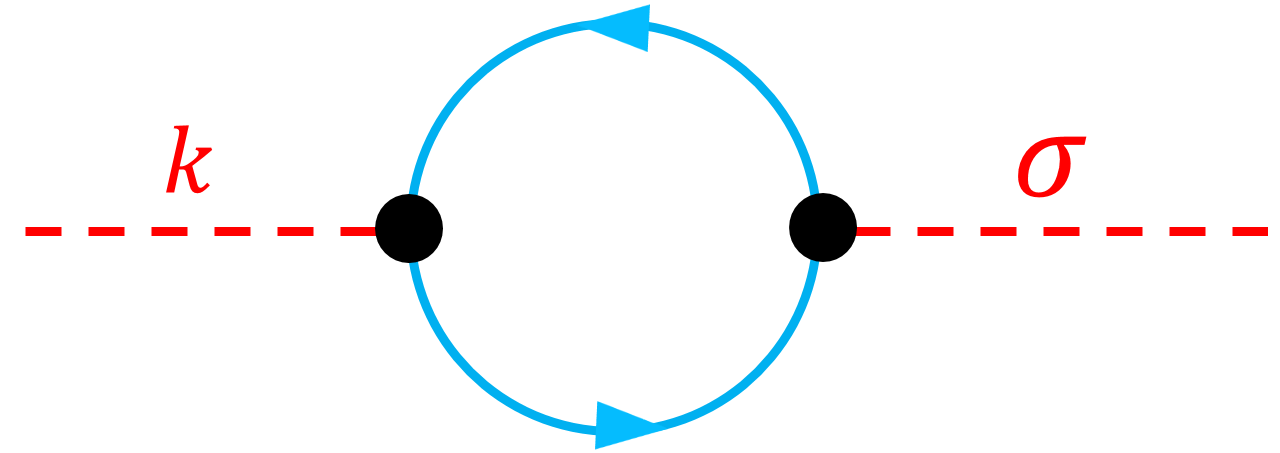}}~~&=-\int\frac{d^4p}{(2\pi)^4}\Tr\left[i S(p)iS(k+p)\right]\nonumber\\&=-\int\frac{d^4p}{(2\pi)^4}\Tr\left[\frac{1}{\slashed{p}-\Delta-\kappa\gamma^5\gamma^0}~\frac{1}{\slashed{p}+\slashed{k}-\Delta-\kappa\gamma^5\gamma^0}\right]~,~~~~~~~
\end{align}
Integrate out $p_0$ first and set a cutoff for the spatial momentum integral, we obtain
\begin{align}
	\mathcal{Z}_\Delta^{00}=&\frac{\kappa^2}{12\pi^2\Delta^2}\left(1-\frac{3\Delta^2}{\kappa^2}\ln \frac{\Delta}{\Lambda_{\sigma}}\right)~,\\
	\mathcal{Z}_\Delta^{33}=&-\frac{\kappa^2}{12\pi^2\Delta^2}\left[\frac{1}{3}-\frac{3\Delta^2}{\kappa^2}\left(\ln \frac{\Delta}{{\Lambda}_{\sigma}}+\frac{1}{3}\right)+\frac{2\Delta^4}{\kappa^3\sqrt{\Delta^2+\kappa^2}}\rm{Arctanh}\frac{\kappa}{\sqrt{\Delta^2+\kappa^2}} \right]~.	
\end{align}
The field strength renormalization coefficient $\mathcal{Z}^{\mu\nu}_{\Theta}$ can be computed analogously. Now the corresponding 1-loop integral is 
\begin{align}
	\parbox{0.25\textwidth}{\includegraphics[width=0.25\textwidth]{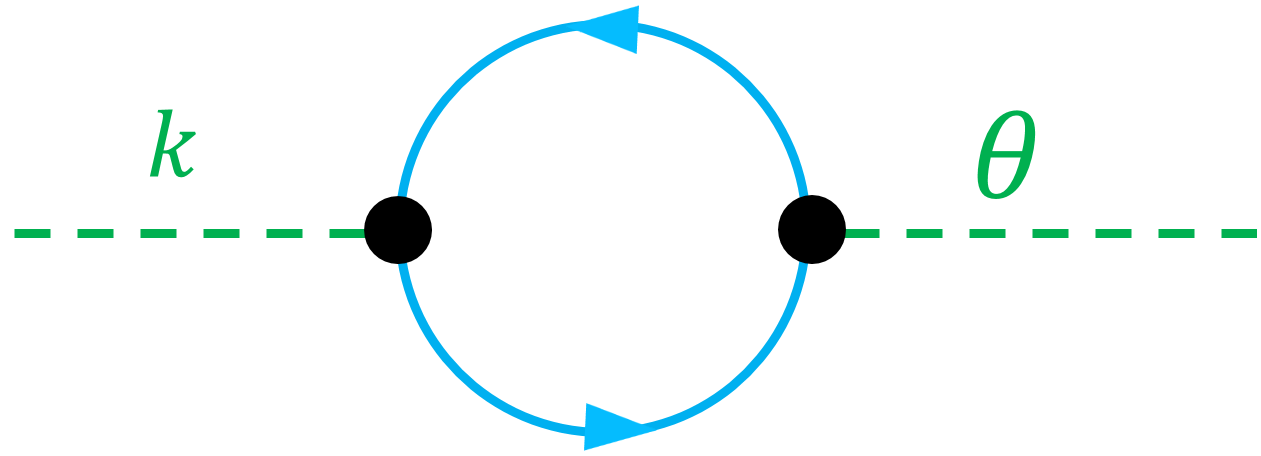}}~~&=-\Delta^2\int\frac{d^4p}{(2\pi)^4}\Tr\left[\gamma^5 S(p)\gamma^5 S(k+p)\right]\nonumber\\&=\Delta^2\int\frac{d^4p}{(2\pi)^4}\Tr\left[\gamma^5\frac{1}{\slashed{p}-\Delta-\kappa\gamma^5\gamma^0}\gamma^5~\frac{1}{\slashed{p}+\slashed{k}-\Delta-\kappa\gamma^5\gamma^0}\right]~,
\end{align}
Performing the momentum integral, the coefficients read 
\begin{align}
	\mathcal{Z}^{00}_\Theta=\frac{\kappa^2}{4\pi^2}\left[1-\frac{\Delta^2}{\kappa^2}\ln \frac{\Delta}{\Lambda_{\theta}}\right],\qquad \mathcal{Z}^{33}_\Theta=-\frac{\kappa^2}{4\pi^2}\left[\frac{1}{3}-\frac{\Delta^2}{\kappa^2}\left(\ln\frac{\Delta}{\Lambda_{\theta}}-\frac{1}{3}\right)\right]~.
\end{align}

\bibliographystyle{utphys}
\bibliography{ref}

\end{document}